\documentclass[12pt,preprint]{emulateapj}

\lefthead{Majewski, Zasowski, \& Nidever}
\righthead{Lifting the Dust Veil}

\slugcomment{Accepted to ApJ: 25 May 2011}

\shorttitle{Dereddening with Near- and Mid-Infrared Photometry}

\shortauthors{Majewski, Zasowski, \& Nidever}

\begin{document}

\title{Lifting the Dusty Veil With Near- and Mid-Infrared Photometry: \\
I. Description and Applications of the Rayleigh-Jeans Color Excess Method}

\author{
Steven R. Majewski, 
Gail Zasowski, \& David L. Nidever}
\affil{Department of Astronomy, University of Virginia,
    Charlottesville, VA 22904}
\email{srm4n, gz2n, dln5q@virginia.edu}

\begin{abstract}

The Milky Way (MW) remains a primary laboratory for understanding the structure 
and evolution of spiral galaxies, but typically
we are denied clear views of MW 
stellar populations at low Galactic latitudes because of extinction by interstellar dust.
However, the combination of 2MASS near-infrared (NIR) and {\it Spitzer}-IRAC mid-infrared (MIR)
photometry enables a powerful  
method for determining the line of sight reddening to any star:
the sampled wavelengths lie in 
the Rayleigh-Jeans part of the spectral energy distribution of most stars, where, to first
order, all stars have essentially the same intrinsic color.
Thus, changes in stellar NIR-MIR colors due to interstellar reddening are readily apparent, 
and (under an assumed extinction law)
the observed colors and magnitudes of stars can be easily and accurately
restored to their intrinsic values, greatly increasing their usefulness for 
Galactic structure studies.
In this paper we explore this ``Rayleigh-Jeans Color Excess'' (RJCE) method 
and demonstrate that use of even a simple variant of the RJCE method based on a single reference color,
($H-$[4.5$\mu$]), can rather accurately remove dust effects from 
previously uninterpretable
2MASS color-magnitude diagrams of stars in fields along the heavily reddened Galactic
mid-plane, with results far superior to those derived from application of other dereddening methods.
We also show that ``total'' Galactic midplane extinction looks rather different from that
predicted using 100$\mu$ emission maps from the IRAS/ISSA and COBE/DIRBE
instruments as presented
by Schlegel et al.
Instead, the Galactic mid-plane extinction strongly resembles
the distribution of $^{13}$CO ($J$=1$\rightarrow$0) emission.
Future papers will focus on refining the RJCE method and applying the technique to understand better
not only dust and its distribution, but the distribution of stars intermixed with the dust in the low-latitude Galaxy.

\end{abstract}


\section{Introduction} \label{sec:intro}

\subsection{The Impact of Dust on Galactic Structure and Stellar Population Studies} \label{sec:impact-dust}

As the only major galaxy for which we can obtain detailed information on the chemistry,
kinematics, and spatial distribution of large numbers of individual stars, the Milky Way (MW)
remains a primary laboratory for understanding the structure and evolution of spiral galaxies.
An important tool in the study of Galactic structure is
the color-magnitude diagram (CMD).  The CMD for stars along a given line
of sight contains information on their distributions of spatial density,
age, and metallicity.
The advent of high precision, linear, wide-area photometry from
electronic array detectors,
aided by the interpretive insights provided by computer models to project the
convolution of population age, metallicity and density distribution into synthesized
stellar color-magnitude
distributions (e.g., Bahcall \& Soneira 1980,
Robin \& Creze 1986, 
Yoshii et al.\ 1987,
Reid \& Majewski 1993, 
Ng et al.\ 1995,
Castellani et al.\ 2002, 
Robin et al.\ 2003, 
Girardi et al.\ 2005), 
is finally allowing authentic realizations of the type of systematic surveying of MW 
stellar populations first attempted by Herschel (1785), and
established as a formal goal of Galactic astronomers more than a century
ago (Kapteyn 1906).  
Profound fulfillment of this long-sought goal --- the production of stellar CMDs having extraordinary statistics
over extensive angular coverage --- is being provided by large-area surveys like the Two Micron All-Sky Survey 
(2MASS, Skrutskie et al.\ 2006),
the Sloan Digital Sky Survey (SDSS, York et al.\ 2000), 
and UKIDSS (Lucas et al.\ 2008), and will be expected to continue with future endeavors
like PAN-STARRS (Kaiser et al.\ 2002), 
the Large Synoptic Survey Telescope (Ivezic et al.\ 2008),
and Sky-Mapper (Keller et al.\ 2007).

However, a fundamental and unresolved difficulty with systematic explorations of the Galaxy,
despite the existence of huge databases of accurate photometry for billions of stars,  
is extinction by interstellar dust.  This obstacle has thwarted astronomers dating back to Herschel,
Struve (1847), and Kapteyn (1909), and, despite having been initially quantified by Trumpler (1930) 80 years ago,
remains a roadblock to our detailed understanding of the inner
MW and to the interpretation of CMDs of its stars.
While the reduced effects of dust on near infrared (NIR)
photometry 
have allowed us to take CMD-probing of the Galaxy to
dustier regions at lower latitudes, 
our view of the densest, but most important, parts of our galaxy is still
challenged by heavy and non-uniform reddening and extinction, which typically renders low latitude field
stellar CMDs virtually uninterpretable in a stellar 
populations context. 
The problem is multi-faceted: (1) 
when beyond modest levels of extinction (when, e.g., $E[B$$-$$V]$ $\gtrsim$ $0.5$), the dust density 
is generally 
variable on angular scales smaller than the resolution of available reddening maps.  
(2) Existing
reddening maps (e.g., Burstein \& Heiles 1982; Schlegel, Finkbeiner \& Davis 1998, ``SFD'' hereafter) 
are {\it two dimensional} --- they give only the {\it total}
Galactic extinction along a line of sight, not just that which is foreground to a
particular star.  At optical wavelengths, star-by-star dereddening from combined
{\it UBV} colors is commonly attempted, but this technique is hampered in
highly extinguished 
regions, where (3) the ``standard Galactic extinction law"\footnote{
For example, characterized by the optical total-to-selective extinction ratio
 $R_V = A_V/E(B-V) = 3.1$ (e.g., Savage \& Mathis 1979, Rieke \& Lebofsky 1985)}
no longer applies (e.g., Cardelli, Clayton \& Mathis 1989), 
and where (4) the extinction pushes
stars to substantially fainter magnitudes (exacerbated by the fact that in such regions $R_V$ can exceed 5).
Finally, (5) in many instances the process of dereddening
the observed colors of a star back to their (unknown) intrinsic values is degenerate because 
the reddening vectors and the distribution of stars in color-color space are either parallel (see \S2)
or intersect at multiple points (e.g., as in the case of {\it UBV} dereddening).

\begin{figure} 
\begin{center}
\includegraphics[angle=90,width=0.45\textwidth]{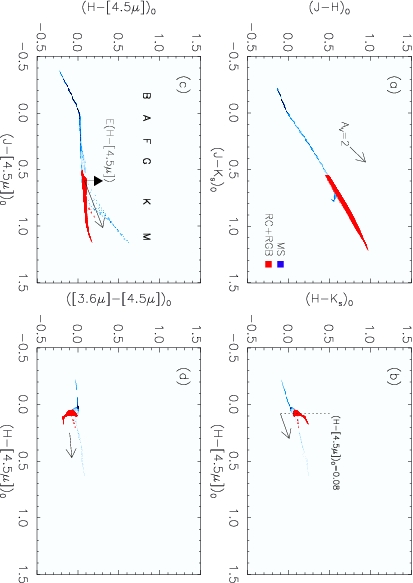}
\end{center}
\caption{Various color-color diagrams showing the ranges in color for a broad sampling of the
Girardi et al.\ (2002) isochrone set (7 $\leq$ {\it log(age/yr)} $\leq$ 10.15, -1.5 $\leq$ [Fe/H] $\leq$ 0.18).   
The dwarf star sequence is shown in blue while giant stars --- i.e. stars in the red clump or along the 
red giant branch --- are overplotted in red.
Reddening vectors (using the Indebetouw et al.\ 2005 extinction curve)
equivalent to two magnitudes of $V$-band extinction are shown in each panel.
The vertical arrow in panel (c) shows the equivalent $E(H-[4.5\mu])$ for this extinction, and 
we indicate spectral types corresponding to $(J-[4.5\mu])_0$ colors for solar metallicity stars.
The axis scales in each panel have been deliberately made identical for the purpose of highlighting the
varying range of values spanned by each color.}
\label{fig:2cd}
\end{figure}

Some of these problems are lessened by moving to near-infrared wavelengths.
For example, 
variations in the extinction law are typically 
considered to be less of a 
problem in the NIR (Cardelli et al.\ 1989,
Indebetouw et al.\ 2005;  although we show elsewhere 
--- Zasowski et al.\ 2009, ``Paper II'' hereafter ---
that there do exist variations in the near- and mid-infrared extinction law that appear to 
be related to Galactic gradients in dust grain size or composition).
On the other hand, 
NIR photometry 
is confounded by an additional problem:
{\it The reddening vector for NIR colors
is almost exactly parallel to the stellar locus in color-color distributions.}  Figure~\ref{fig:2cd}a
demonstrates the problem for NIR two-color combinations: the process
of correcting the observed {\it JHK$_s$} colors of a star back to the intrinsic
values is severely degenerate 
between reddening and intrinsic stellar type
(particularly when both intrinsic and observational 
color scatter of stars along these loci
is considered).  The problem is the same for all
permutations of broadband NIR colors.

However, as we shall demonstrate,
the combination of 2MASS and {\it Spitzer}-IRAC 
photometry (or any near- and mid-infrared dataset)
allows a direct and reliable assessment of the line of sight
reddening to any particular star --- and across wavelengths where the
reddening law is more universal (Rieke \& Lebofsky 1985, Cardelli et al.\ 1989 ---
although not entirely so,  Paper II).
Moreover, at these wavelengths,
the color effects of reddening and stellar atmospheres are almost completely
separable, an advantage that is the foundation of the present series of papers (see \S\ref{sec:rjce-context}).

\subsection{Color Excess as a Tool for Measuring Extinction} \label{sec:color-excess}

With Trumpler's (1930) determination that the observed brightnesses of stars are modulated by the
presence of interstellar dust, a definitive explanation of the previously known 
(e.g., \"Opik 1929, Gerasimovi\v{c} 1929) 
but not understood phenomenon of stellar reddening
--- whereby a star has a measured color that is redder than expected for its spectral type ---
was at hand.  
\"Opik (1931) explored several hypotheses for the origin of reddening (as seen in early-type stars)
and concluded that absorption of light by interstellar dust clouds was the most probable 
explanation.\footnote{\"Opik (1931) laid out the problem 
of ``cases of a general inconsistency between color and 
spectrum'' and explored several hypotheses --- both ``stellar'' (atmospheric or circumstellar)
and ``interstellar'' in nature --- for its origin (as seen in early type stars); 
though finding apparently workable
alternative hypotheses to explain the creation of  color excesses, he 
cited as most probable ``absorption by dust clouds in space, consisting of particles of various
sizes, with a `meteoric' distribution of the diameters of the particles  (see also Elvey 1931 and
Elvey \& Mehlin 1932). \"Opik's countrymen, 
Wallenquist (1929) and Schal\'en (1931), apparently were already convinced
that selective absorption of starlight caused color excesses.  
The conclusions of these Swedish researchers were arrived at primarily by 
study of the expected spectral variations on stellar spectral energy distributions
exacted by the various proposed phenomena.  Trumpler's 1930 results, 
though circumstantial, provided strong and satisfying
confirmation that the contemporary theories of the spectral effects of scattering by interstellar particles
were on the right track.  Additional evidence was provided by the recognition that there was
a correlation between the intensity of interstellar lines and color excesses.}

Soon an association of at least some of this selective absorption with dark clouds (of the type
photographed extensively and famously by E.~E.~Barnard and S.~I.~Bailey decades before) 
was made through what may be the first published color excess (i.e., reddening) 
maps across an individual cloud (one associated with the open cluster NGC 663) 
by Carol Jane Anger (1931), and then across
a general region (Cepheus) by Elvey \& Mehlin (1932).  These studies
were soon followed by the first rudimentary large-scale maps of optical color excess as a 
function of longitude, latitude, and distance (e.g., Elvey 1931, Zug 1933, Stebbins 1933, Stebbins
\& Whitford 1936), 
which immediately revealed the global non-uniformity of the distribution of the reddening particles.
Williams (1934) summarizes much of this early pioneering work.
Nevertheless, for decades 
researchers have been frequently
tempted or forced to use simple assumptions about the distribution of dust --- e.g.,
the layered models giving rise to
cosecant{($b$)} laws, and/or spatial homogeneity leading to assumed
linear relations of color excess with distance.\footnote{It is noteworthy 
that the {\it ubiquity} of interstellar absorption was quickly recognized, with the finding of color excesses
in stars as far away as 300 pc at the Galactic poles (e.g., Stebbins, Huffer \& Whitford 1939);
this alone attested to a rather thick distribution of dust about the Galactic plane.
It is also telling that at least some scientists proceeded with great trepidation while making 
critical, simplifying assumptions of the type that are often still made even today: 
``If the interstellar material is not thoroughly `stirred' so to speak, 
and the ratio of total to selective absorption is not reasonably constant over 
large regions, then the problem of determining absorption and distances in the galaxy
from measures of color is of course hopeless.'' (Stebbins, Huffer \& Whitford 1939)  }

A correlation between the color excess and the total space absorption was subsequently
identified, and soon the ratio between the selective and total extinction was
quantified for different optical wavelengths (e.g., Stebbins \& Whitford 1936, Greenstein \& Henyey 1941),
which led to the first attempts at defining the extinction law with wavelength
(Stebbins, Huffer \& Whitford 1939).  With a reliable extinction law, observed
color excesses could be translated back into total extinctions, a necessary
step in mapping the distribution of both dust and stars in the Galaxy.
Although alternative methods have been developed to measure line-of-sight 
extinction (e.g., through counts of background galaxies, or by mapping 
proxies for dust obscuration, such as the intensity of radiation by associated
gas or thermal emission by the dust itself), use of color excesses have always
provided the most reliable means by which to measure extinction along the
line of sight to a star, and, specifically, {\it foreground} to that star.

\subsection{Early Infrared Color Excesses and Spectroscopy} \label{sec:early-ir}

The use of 
infrared color excesses to calculate levels of dust extinction,
taking advantage of the reduced influence of extinction at longer wavelengths, 
has a history primarily in the exploration and mapping of the extinction
in dark (i.e., molecular) clouds.  
For example, Jones et al.\ (1980) explored the observed distribution of 
$(J-H)$ and $(H-K)$ colors of stars in the direction of a specific Bok globule in
the Coalsack Nebula to estimate reddenings.

As shown in Figure~\ref{fig:2cd}, the problem with this sort of analysis is the degeneracy
between the NIR reddening vector and the stellar locus for most spectral types, which leaves
as generally inseparable
the intrinsic colors of the stars versus the degree of reddening to each star.
Nevertheless,  
with infrared spectra for only a small fraction of their 75
tracer stars available to constrain the general types of stars in their photometric survey,
Jones et al.\ estimated the radial variation in extinction (derived 
through $E[J-K]$) in the globule 
after assuming all stars to be of a specific 
spectral type (M3 III).
A similar approach was applied by Hyland
(1981) to map six Bok globules, by Casali (1986) for another six, and by Smith (1987)
for the Carina nebula.

While follow-up study of their Bok globule by Jones et al.\ (1984) 
increased to several dozen
the number of available spectra (along with other data) for classifying
the stars and improving the reddening estimates, this effort only
highlighted the labor-intensive
nature of pairing photometry with spectroscopy on a star-by-star basis to
break the degeneracies between intrinsic NIR stellar colors and the reddening
vectors.  
More recently, Arce \& Goodman (1999b) compare and summarize the quality
various techniques for mapping extinction, and assert that the ``most exact method for measuring
extinction [is] using the color excesses of individual stars with measured spectral type.'' 
However, as they point out, ``the real drawback of this technique is the large amount of
time required to measure each and every star's spectrum.''
A reliable means by which to estimate stellar type and reddening
{\it from photometry alone} would clearly
allow much more flexibility and the ability to make and utilize large photometric surveys.

\subsection{Statistical Color Excesses in the Near-Infrared: NICE, NICER and Related Methods} \label{sec:nice-r}

The early dark cloud
studies listed in \S\ref{sec:early-ir}, which undertook a ``statistical'' approach to deriving color excesses 
--- i.e., basically assuming an ``average'' intrinsic color for sources against which to 
estimate color excesses --- were relatively successful
because of their use of infrared data in highly reddened regions: the actual range
of intrinsic NIR colors of the stars was relatively narrow 
compared to the degree of reddening being mapped.

An important development along this path of statistical IR reddening mapping was made by
Lada et al.\ (1994), who, in their extensive study of the molecular cloud IC5146, 
adopted an improved statistical approach to determining 
color excesses that they called the Near-Infrared Color Excess (NICE) method.  
By focusing solely on stellar $(H-K)$ colors, Lada et al.\ (1994) were
able to take advantage of the 
even more limited intrinsic stellar color range (assumed 
by them as $0 < (H-K)_0 < 0.3$ for stars with spectral types from A0 to late M ---
see also Figure~\ref{fig:colhist} below)
to substantially reduce uncertainty in derived NIR reddening values for random field
stars.  These authors point out 
that even simply assuming that the average intrinsic color of field stars is 
$(H-K)_0 = 0.15$ leads to a maximum uncertainty in the ultimately derived 
total extinction, $A_V$, of less than 2.5 magnitudes.
However, while residual uncertainties in the extinction
at the level of 2.5 mag may be sufficiently negligible in the study of dense molecular
clouds reaching extinctions of $A_V$ = 40 mag, for mapping the more diffuse 
extinction of the typical interstellar medium,
the large reddening-to-extinction conversion factor leads to severe fractional $A_V$ errors for a given $E(H-K)$ error.

It is possible to reduce uncertainties by averaging over many stars,
but this can severely limit 
the resolution of the final extinction map.
For example, Lada et al.\ (1994) spend considerable effort modeling their
results to account for cloud structure fluctuations on scales smaller than the resolution
of their maps (which had 1.5 $\times$ 1.5 arcmin cells; see also Alves et al.\ 1998, 
Lada, Alves \& Lada 1999, and Arce \& Goodman 1999b).  Other disadvantages
of the method (at least from the standpoint of understanding individual clouds) include the
difficulty of understanding the relative placement of the stars along the line of sight, the potential for
certain stars to suffer circumstellar extinction (e.g., dust shells), the varying density of stars on the sky, 
the non-constancy of stellar populations with position in the sky
(see especially Arce \& Goodman 1999b and Popowski et al. 2003 on this point),
the dilution of high extinction peaks because of averaging,
and the requirement that the dispersion in the
mean colors of stars in the control fields be significantly smaller than in the target field,
since the latter serves as a limit on the ultimate uncertainties.  All
of these affect the usefulness of applying the NICE method more broadly along the Galactic plane
 (Alves et al.\ 1998, Froebrich \& del Burgo 2006).\footnote{Nevertheless, 
 the method has been applied to create a large
 area map of the Galactic anticenter with 4 arcmin resolution 
 using 2MASS photometry by Froebrich et al.\ (2007), as well as all-sky maps by 
 Rowles \& Froebrich (2009).   Dobashi et al.\ (2008, 2009) also used the method for their maps of
 dust extinction in the Large and Small Magellanic Clouds.} 
Clearly, it would be much better
to know the intrinsic color of {\it each} star to maximize the resolution of reddening
maps to the scale of star-to-star spacing.

Lombardi \& Alves (2001) set out to generalize the Lada et al.\ (1994) NICE method to take advantage of
multi-band data (specifically, {\it JHK$_s$} data from 2MASS) 
and introduced the NICER (Near-Infrared Color Excess -- Revisited) technique.  
In their specific application, Lombardi \& Alves use both $(J-H)$ and $(H-K_s)$ colors 
and the NICER technique to make extinction maps over a large area (625 deg$^2$) around
the Orion and Mon R2 region.
In principle, the main problem with including $J$-band photometry in these color
excess methods is the larger scatter in intrinsic $(J-H)_0$ colors for stars, which is
more than three times larger than in $(H-K_s)_0$ (Girardi et al.\ 2002; 
see also Figures~\ref{fig:colhist}b and \ref{fig:colhist-tri}b below).  
However, Lombari \& Alves argue that
the complication of including the larger intrinsic scatter brought by incorporating
$(J-H)$ colors in the analysis is offset 
by the typically smaller 
uncertainties one encounters
in $J$ band photometry compared to $K_s$, as well as the smaller numerical 
ratio of $A_V/E(J-H)$ compared to $A_V/E(J-K_s)$  (they used 9.35 versus 15.87; Rieke \& Lebofsky 1985), 
which
diminishes the effect of color excess uncertainties once converted to total
extinctions.  The NICER method seeks to balance these trade-offs between
$E(J-H)$ and $E(H-K_s)$ derived maps with an optimization
via maximum likelihood; however, the method is still intrinsically based on
taking advantage of {\it average} field star colors, assuming constancy of that mean
color with position in the sky, and spatially smoothing over 
numerous stars to create each ``pixel'' in a map.  Thus, for example, the Lombardi \& Alves
map of the Orion cloud yielded some regions with ``negative extinction'' owing to 
the presence of many contaminating blue stars from the open cluster NGC 2204.  Nevertheless,
Lombardi \& Alves show that for the same angular resolution,
the NICER method yields less noisy extinction maps that are more sensitive
to lower extinction than the NICE method using $(H-K_s)$ colors alone.
This group has successfully applied the same methodology to explore
the Pipe Nebula (Lombardi, Alves \& Lada 2006) and the Ophiuchus and Lupus Cloud
Complexes (Lombardi, Lada \& Alves 2008).  
More recently, Gosling et al.\ (2009) have updated the NICER method 
to include a variable extinction law; they apply their ``V-NICE'' method 
to understanding the nuclear bulge,
assuming as their baseline intrinsic colors some mean 
$(J-H)$ and $(H-K_s)$ colors of G, K, and M stars.

\subsection{The Rayleigh-Jeans Color Excess Method (RJCE) In Context} \label{sec:rjce-context}
As implied in \S\ref{sec:impact-dust}, a primary motivation for our investigation of improved dereddening methods
derives from an interest in stellar populations and Galactic structure, a fact that shapes our 
dereddening strategy and desired data products.  To these ends we have developed 
the Rayleigh-Jeans Color Excess (RJCE\footnote{For convenience, the
authors have adopted the shorthand pronunciation of this acronym as ``rice''($\backslash$ra{\tiny I}s$\backslash$).}) Method, 
which builds on and contrasts with previous reddening studies in the following ways:

\begin{itemize}

\item As with numerous investigations 
stretching back to the 1930s (\S\ref{sec:color-excess}), we use stellar color excesses to 
gauge the amount of foreground dust.

\item Like NICE, NICER, V-NICE and related studies (\S\ref{sec:nice-r}), 
we rely on {\it infrared} colors to measure the
excesses because these can be gauged more reliably due to the narrower intrinsic color range
of stars at these wavelengths.
This reason is of course in addition to the advantageous facts that infrared extinction is both weaker than optical
extinction (which allows greater distances to be explored in dusty regions) 
and the extinction law is less variable with wavelength and 
environment than that at shorter wavelengths.

\item Similar to the Imara \& Blitz (2006), 
Froebrich et al. (2007), Rowles \& Froebrich (2009), Dobashi et al. (2008, 2009), and Gosling
et al. (2009) studies (\S\ref{sec:nice-r}), 
we are interested in {\it large-scale mapping} of dust (reddening) around the Galaxy ---
i.e., not just that around specific clouds, but also down to rather low dust column densities where
a high precision method is required to assess the color excesses.

\item However, because the stars themselves are a primary interest, 
we are not content with simply assuming their intrinsic photometric 
properties as in most of the large-scale, statistical studies; 
rather, we want to {\it measure} these properties.
The stars are not just 
a means to an end (i.e., evaluating and 
understanding the dust) to be averaged over; rather, the
distribution and properties of the stars are end goals in themselves.
By accurately reproducing the intrinsic color-magnitude distributions of the stars,
we can learn about the distribution of stellar populations within the Galaxy (with a particular
emphasis on the much more obscured and less well-studied populations
at low latitudes), as well as the density laws of the dusty inner Milky Way
via photometric isolation of various ``standard candles.''
As a bonus, with accurate knowledge of the 3-dimensional distribution of
stellar types, we can in turn hope to map the 3-dimensional distribution of the dust itself.

\item Because we seek an accurate means to attain accurate {\it star-by-star foreground reddening}
(as opposed to statistically-assessed, area-averaged, line-of-sight reddening), the
desired method to deal with extinction must gauge the reddening to each star and,
at the same time, recover that star's intrinsic photometric properties.

\item The RJCE method we describe meets this requirement, and we believe it is
the first technique to utilize a combination of 
near-infrared (NIR) {\it and} mid-infrared (MIR) photometry for dereddening on a large scale.  
By exploiting these combined wavelengths,
he color effects of reddening and stellar atmospheres can be almost completely
separated: 
The {\it observed} NIR-MIR colors contain information on the foreground reddening to
a star {\it explicitly}, whereas the NIR-only SEDs (especially those using the $J$-band filter)
contain information on the stellar types.  Again, previous extinction studies using solely NIR
photometry cannot achieve this separability because of the degeneracy
of the reddening vector and the intrinsic distribution of stellar NIR colors.

\item Unlike previous studies, the
RJCE methodology includes both an accurate, star-by-star dereddening procedure as well 
as a second step of recovering and exploiting the recovered intrinsic stellar spectral types for
further improvements in our understanding of the distributions of both the stars and dust.
\end{itemize}

\section{The Rayleigh-Jeans Color Excess Method (RJCE)}

\subsection{Foundation and Procedure}
\label{sec:howitworks}
\begin{figure} 
\begin{center}
\includegraphics[angle=90,width=0.47\textwidth]{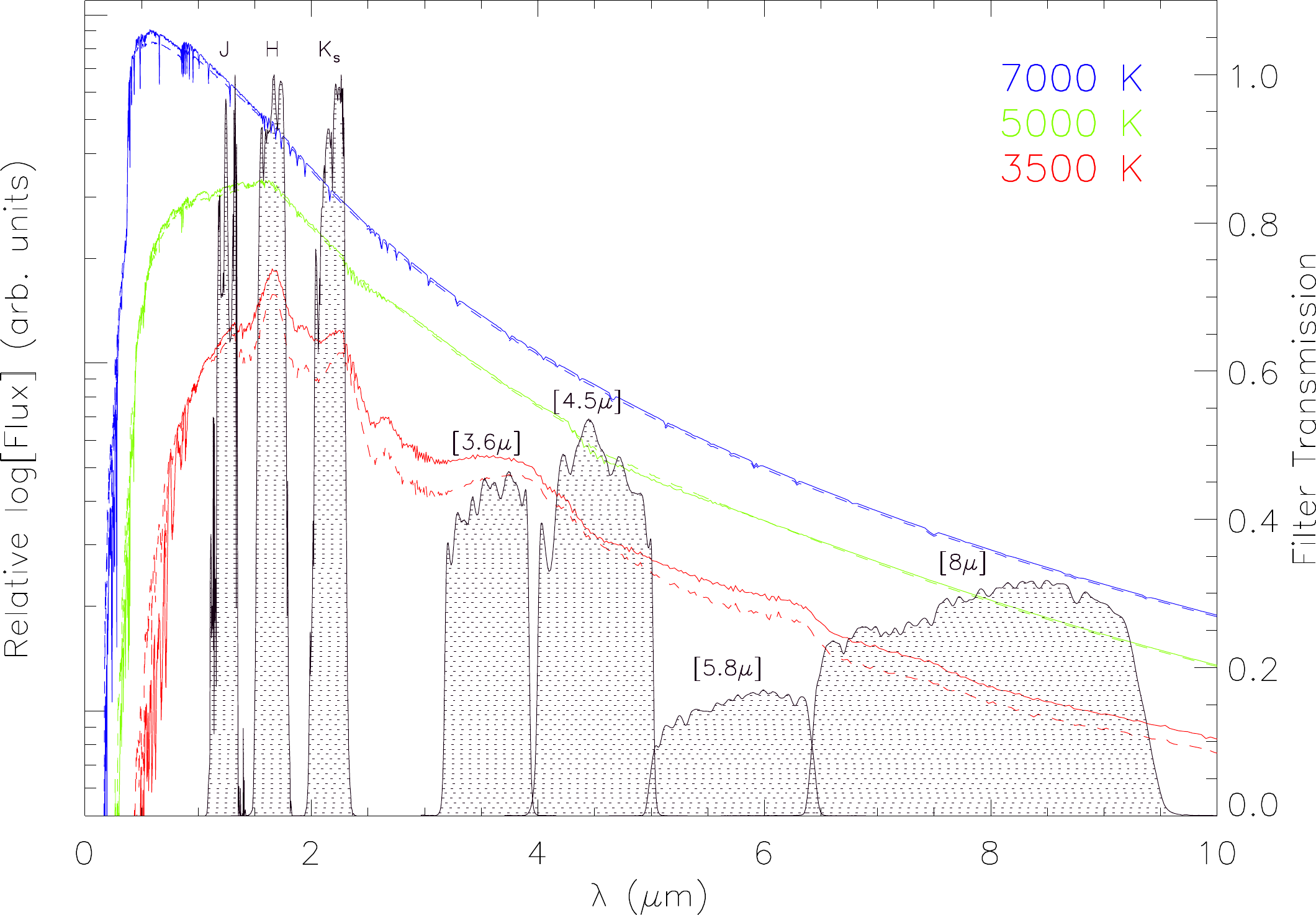}
\end{center}
\caption{Castelli \& Kurucz (2004) model
stellar spectral energy distributions (SEDs) for
stars of three effective temperatures (7000, 5000, and 3500 K) against the bandpass
transmission curves for the 2MASS and {\it Spitzer}-IRAC filters.
The Castelli \& Kurucz SEDs shown represent giant stars 
in the wide F--M spectral range 
with [Fe/H] $\sim$ -1 (dashed lines) and 0 (solid lines).
Note the strong similarity in SED shapes across the MIR filters compared
to the variation across the NIR filters (particularly in the $J$-band), a contrast in behavior that
together forms the basis of the RJCE method.}
\label{fig:models}
\end{figure}

The dereddening step of the RJCE methodology builds upon the 
advances achieved with the NICE method --- and, moreover, allows us to take 
color excess dereddening from the realm
of a statistical to a star-by-star procedure --- because of two factors:
First, the degeneracy between NIR stellar colors and reddening vectors 
is virtually eliminated as one moves to longer wavelength photometry.  This is 
evident, for example, through comparison of 
the reddening vector and stellar locus for the most relevant spectral types
shown in Figures~\ref{fig:2cd}c and \ref{fig:2cd}d.\footnote{While it is true that the reddening vector
 is parallel to the stellar locus in all of the panels of Figure~\ref{fig:2cd}
 for K and M type dwarfs, these are actually extremely rare stellar types in 
 large infrared surveys like 2MASS and GLIMPSE, due to their intrinsic faintness.
 This can be seen later in Figure~\ref{fig:2cd-tri}, which is the same as Figure~\ref{fig:2cd}
 but for a model Galactic mid-plane field as might be observed with GLIMPSE; that example
 shows no representation by late type dwarfs.}  
But a second key feature that MIR photometry --- such as that afforded by IRAC
on the {\it Spitzer Space Telescope} --- adds to the Galactic structure
toolbox is the measurement of fluxes on what is ostensibly the Rayleigh-Jeans part
of the spectral energy distribution (SED) of most stars.  In the limit that stellar atmospheres radiate as 
perfect blackbodies, their colors are all the same
across the MIR bands for most stellar temperatures
(spectral types)\footnote{This uniform color would nominally be equal to 0 
in the Vega-based 
magnitude system, except that the Vega SED itself departs from that of a blackbody, 
predominantly due to the presence of hydrogen absorption lines.};
this means that observed
departures from this standard color {\it directly and explicitly} reflect
reddening by dust.  
We show below 
that even in the case of real stellar atmospheres, the flux ratios across the various MIR color combinations, 
even including the reddest NIR bands, 
are generally constant enough across stellar types 
(i.e., the color variations are small enough) so as to make the RJCE method remarkably robust.
Figure~\ref{fig:models} shows realistic stellar SEDs (i.e., non-blackbody models) and their uniformity of slope at MIR wavelengths, and
illustrates the degree to which we can expect nearly-identical colors for most normal stars with long-wavelength and long-baseline filter combinations.

Figure~\ref{fig:2cd}c demonstrates two examples of how reddenings
might be measured using the power of combining NIR and MIR photometry:

(1) As has been traditionally done, 
combinations of measured colors can be used to determine the extinction to a star by 
dereddening along color excess vectors projected back to the
intrinsic stellar locus.  As mentioned above, reddening vectors for NIR bands {\it combined}
with MIR bands are generally {\it not} degenerate with the stellar
locus for most stars typically encountered (e.g., compare reddening vectors in 
Figures~\ref{fig:2cd}c and \ref{fig:2cd}d with that in \ref{fig:2cd}a).  
Other combinations of NIR-MIR filters 
than those shown in Figure~\ref{fig:2cd} also show promising degrees of
orthogonality between the reddening vector and stellar locus, and, in theory, one could use the combination of many
NIR-MIR color combinations to determine the reddening to a star.  This procedure would be equivalent
to performing, e.g., a least squares fit to the linear combination of the
infrared extinction law (i.e., scaling all color excesses to a common reddening, say,
$E[J$$-$$K_s]$)
and the SED of the stellar type at
the intersection of the reddening vector and the stellar locus.  As demonstrated by Figure~\ref{fig:2cd}c, for most 
typically-encountered stellar types there is only a single such intersection point.

(2) 
As seen in the particular example in Figure~\ref{fig:2cd}c, the intrinsic $(H-[4.5\mu])_0$ 
variation of  
stellar colors is very small over a wide range of spectral types: less than 0.1 mag across 
F, G and K stars (spanning a very large range of age and [Fe/H])
and only
0.4 mag
across the entire range of B through M spectral type (Girardi et al.\ 2002), ignoring late type
dwarfs in both cases.
The near constancy of ($H-[4.5\mu])_0$ colors means that the foreground interstellar reddening, $E(H-[4.5\mu])$,
can be estimated from the {\it observed} ($H-[4.5\mu]$) {\it alone} for most 
typically-encountered stars; conversion of this particular
color excess to an extinction (we choose to use the $K_s$ band) is possible through the adoption of an IR extinction law
and intrinsic $(H-[4.5\mu])_0$ value:
\begin{equation} \label{equ:rjce}
\displaystyle A(K_s) = 0.918 (H-[4.5\mu]-0.08),
\end{equation}
\noindent 
Obviously, when using this method, the stronger the reddening is, the less significant
are the effects of the small variation in intrinsic stellar colors.
Moreover, in most directions of the Galaxy, including in
the Galactic mid-plane, the vast majority of stars seen in 2MASS CMDs (for example)
are evolved 
F, G and K stars
with colors between those 
of a several-Gyr-old population's main sequence turn-off and red giant branch tip.
We demonstrate the efficacy of this particular, simple, single-color version of RJCE dereddening
below (Figure~\ref{fig:sfd_cmd}),
but investigation of other combinations of colors (some also explored below) show that this procedure
works more or less for any combination of NIR-MIR filters sampling the Rayleigh-Jeans part of SEDs
(e.g., $H$, $K_s$, [3.6$\mu$], [4.5$\mu$], but not $J$),
so that it is possible to make multiple, independent measures of the reddening for each star 
in this way. 

In future papers we plan to
discuss in detail the effects of the (small) intrinsic color variations in
stars, as well as to
optimize methods for taking advantage of multiple color data.  
Here we focus on a couple of example 
variants of the RJCE method and compare
extinction
maps made with RJCE and 2MASS+GLIMPSE-detected
stars to those made using other infrared excess 
dereddening methods (\S\ref{sec:cp-maps}) and dust tracers (\S\ref{sec:cp-othermethods}).
In particular, we show that RJCE-generated extinction maps of the midplane
bear little resemblance to those derived by Schlegel et al.\ (1998) 
from 100$\mu$ emission, while,
on the other hand, they do strongly resemble the distribution of $^{13}$CO ($J$ = 1$\rightarrow$0) molecular
cloud emission, as well as the distribution of overall 2MASS starcounts. 
Through demonstration of these various correlations and non-correlations among the different
maps, we highlight the inadequacy of long wavelength infrared dust {\it emission} to serve as a
suitable linear proxy for dust {\it extinction}.

We also explore the power of the RJCE methodology to recover stellar type information (\S\ref{sec:cp-cmds}), 
discuss several limitations of the RJCE method (\S\ref{sec:limits}), 
and show a few test applications that demonstrate the
potential of the RJCE method to both overcome and explore Galactic dust (\S\ref{sec:ex-apps}).
In a companion paper (Nidever, Zasowski \& Majewski 2011, ``Paper III'' hereafter) we provide 
high resolution, electronic versions of the new, two-dimensional 
extinction maps we have created across the entire GLIMPSE-I 
and Vela-Carina ({\it Spitzer} PID 40791) IRAC survey areas
using the methods outlined in this paper.  In the future we expect to improve the 
quality of these maps as 
we refine the simple version of the RJCE method used here. 
One ultimate goal is to apply this dereddening
methodology to ascertain the 3-dimensional distribution of both dust and stars within the Galactic 
midplane, lifting the dusty veil at low
latitudes to facilitate new approaches to persistent, as well as recently 
recognized, problems in Galactic structure and stellar population studies.

\subsection{Color Options for RJCE} \label{sec:col-options}

\begin{figure}
\begin{center}
\includegraphics[width=0.45\textwidth]{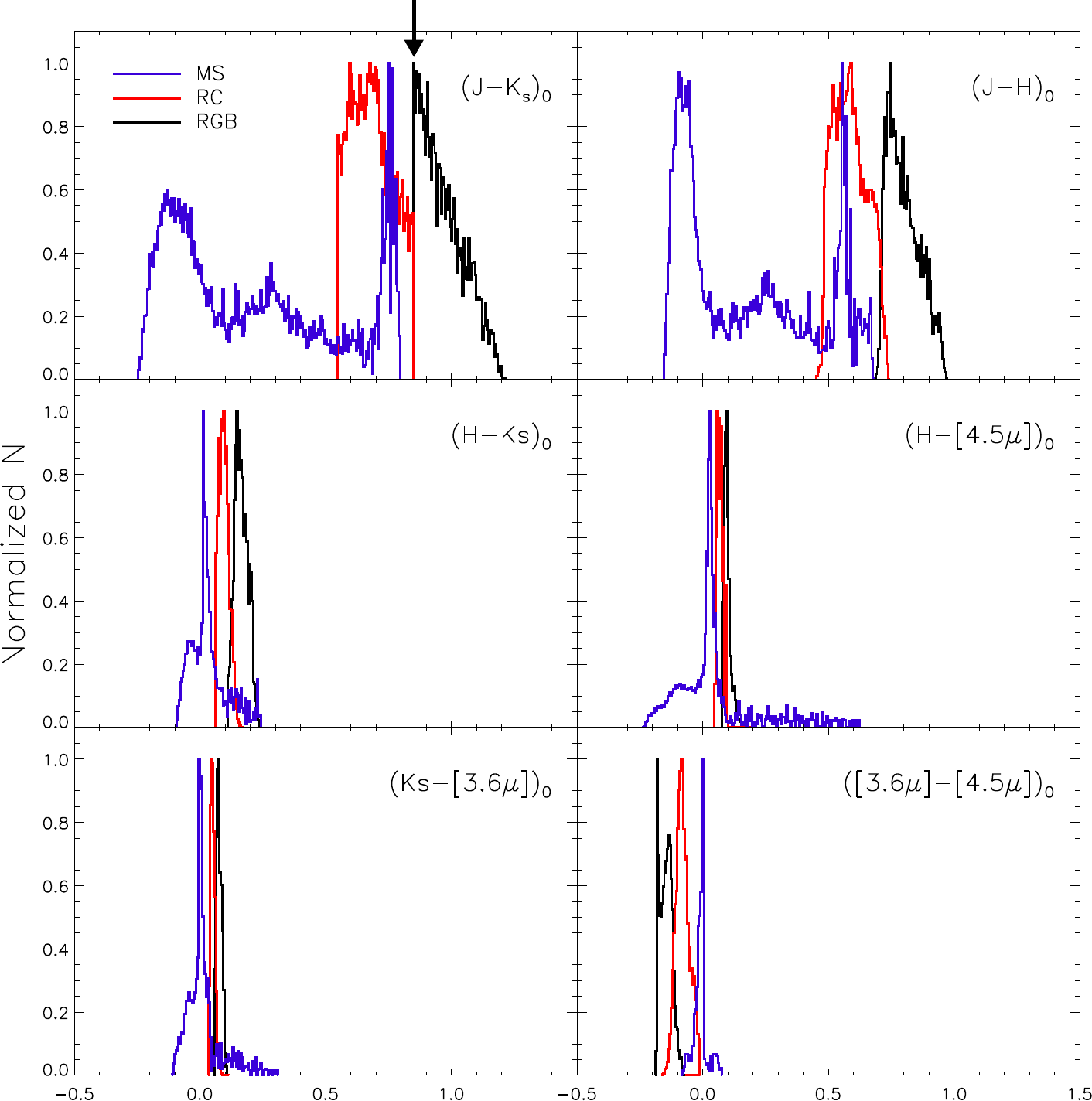}
\end{center}
\caption{Histograms showing the color distributions of selected color combinations from Figure~\ref{fig:2cd}
and weighting each point in the Girardi et al.\ (2002) isochrone set equally within the separately
shown histograms for each stellar type: MS (blue), RC (red), and RGB (black). 
The arrow at $(J-K_s)$ = 0.85 indicates the color chosen to distinguish RC and RGB stellar models.}
\label{fig:colhist}
\end{figure}

In this subsection, we evaluate
some of the NIR-MIR color combinations presently
available for use with the RJCE method, with a particular focus on data from 2MASS and {\it Spitzer}-IRAC.
The histograms in Figure~\ref{fig:colhist} summarize 
the expected stellar color ranges for the filter combinations shown in Figure~\ref{fig:2cd}
and for main sequence (MS), red clump (RC), and red giant branch (RGB) stars separately.
We have divided the evolved stars into the
RC (0.55 $\leq$ $[J-K_s]_0$ $\leq$ 0.85) and RGB (0.85 $\leq$ $[J-K_s]_0$ $\leq$ 1.4) groups by color,
guided by the Girardi et al.\ (2002) isochrones as well 
as TRILEGAL modeling.

In \S\ref{sec:howitworks}, we proposed that colors composed of filters falling on the Rayleigh-Jeans portion of the stellar SED
would have values nearly independent of stellar type.  Figure~\ref{fig:colhist} further explores the degree to which this hypothesis holds true.
Here, the range of intrinsic colors
over all three sampled groups of stars (MS, RC, RGB),
as well as the ranges within each group,
generally decrease with colors made from increasingly redder
filter passbands, though there are some subtleties.  For example, the $(H-[4.5\mu])_0$
color ranges are about as narrow as those for $(K_s-[3.6\mu])_0$, presumably because
the broader wavelength baseline of the former color combination compared to the latter
compensates for the slightly greater sensitivity of $H$ to stellar SED variations than $K_s$.  
Nevertheless, in principle, either of these two
color combinations offer very promising RJCE 
method possibilities because of the negligible range of intrinsic stellar colors.
In this paper, we have generally opted for use of $(H-[4.5\mu])_0$ as our baseline
RJCE method color index. 
Figure~\ref{fig:colhist} also demonstrates how strongly sensitive colors that include the $J$-band are to stellar type/temperature, with
both $(J-H)_0$ and $(J-K_s)_0$ spanning more than a magnitude for normal stars.  It is precisely this sensitivity that we aim to exploit {\it after}
dereddening with $(H-[4.5\mu])$ colors to separate stars by their temperature/type.

\begin{figure} 
\begin{center}
\includegraphics[angle=90,width=0.48\textwidth]{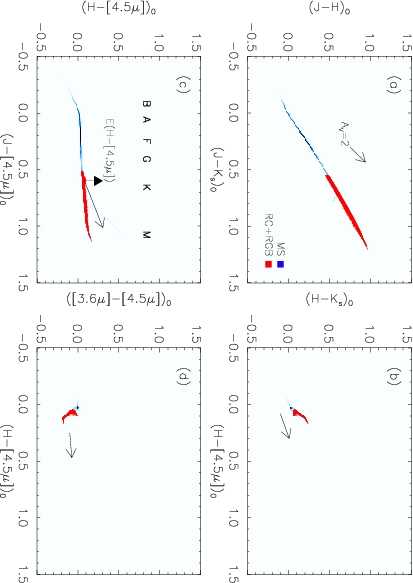}
\end{center}
\caption{Four color-color diagrams showing the ranges in color for 
the TRILEGAL model of a 4.0 deg$^2$ field centered at $[l,b]$ = $[25,0]^{\circ}$ (Girardi et al.\ 2005).  As in Figure~\ref{fig:2cd}, 
the dwarf sequence is shown in blue and the giants (RC+RGB) are overplotted in red.
Reddening vectors (using the Indebetouw et al.\ 2005 extinction law)
equivalent to two magnitudes of $V$-band extinction are shown in each panel.  
The vertical arrow in panel (c) shows the equivalent $E(H-[4.5\mu])$ for this extinction.}
\label{fig:2cd-tri}
\end{figure}

\begin{figure}
\begin{center}
\includegraphics[width=0.45\textwidth]{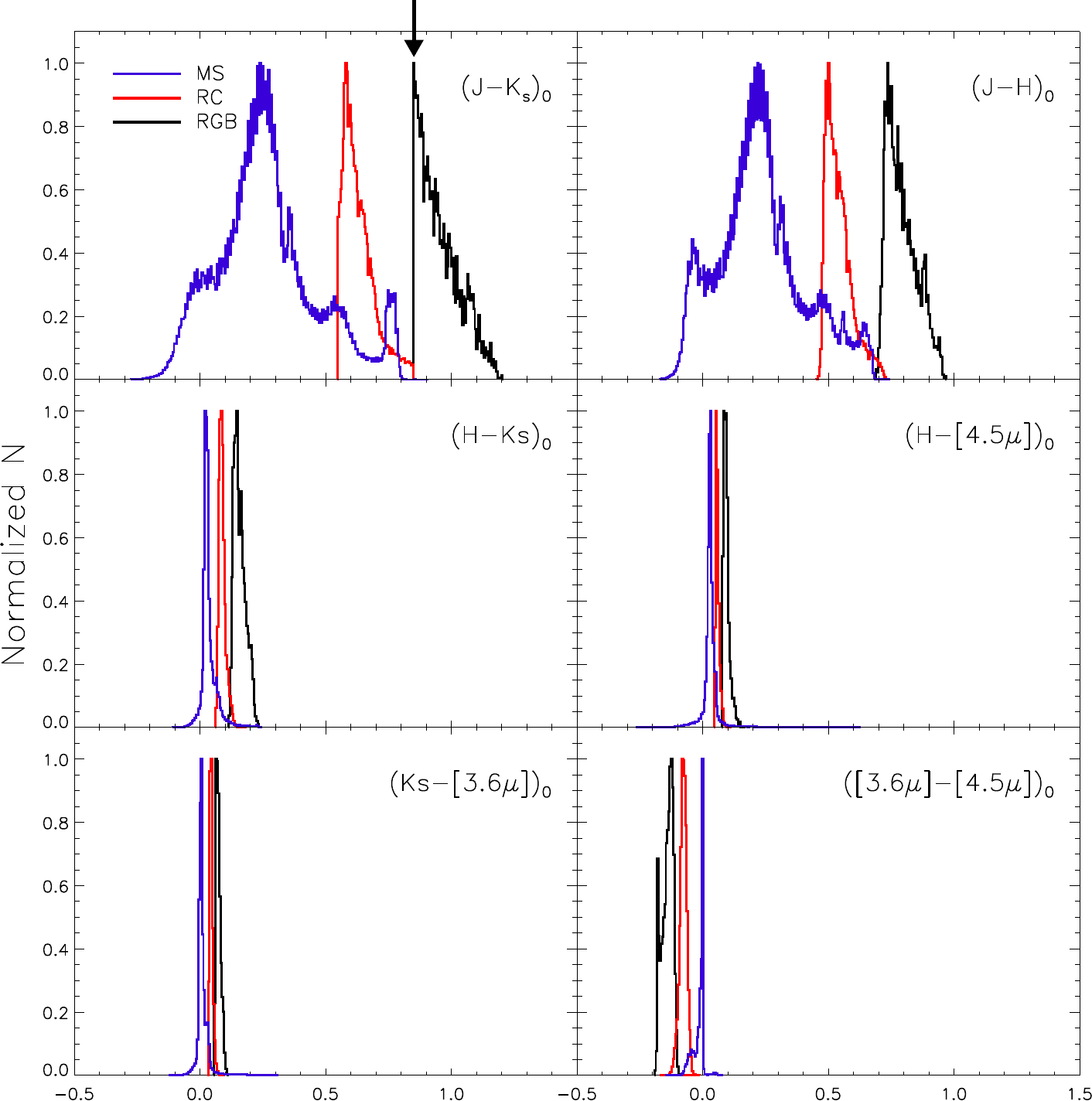}
\end{center}
\caption{Same as Figure~\ref{fig:colhist} but with the stellar colors weighted by the TRILEGAL model data shown in Figure~\ref{fig:2cd-tri}. }
\label{fig:colhist-tri}
\end{figure}

Figures~\ref{fig:2cd} and \ref{fig:colhist} are slightly unphysical because each point in the Girardi et al.\ isochrones
is given equal weight (in Figure~\ref{fig:colhist} the points within each stellar group are given equal weight, 
but each of the histograms is normalized to peak at unity).  In contrast, any particular direction in
the Galaxy will
be weighted towards higher representations of specific age/metallicity populations, and this will
generally narrow the intrinsic color ranges of the actually sampled 
stars even further.  To demonstrate this point 
Figures~\ref{fig:2cd-tri} and \ref{fig:colhist-tri} present analogous distributions to 
Figures~\ref{fig:2cd} and \ref{fig:colhist}, respectively, but
now using stellar distributions derived from a TRILEGAL model of a typical GLIMPSE-like field
(simulating 4.0 deg$^2$ at $(l,b)$ = $(25,0)^\circ$ with zero extinction).
In this case, the weighting in the color distributions are affected by the specific adopted age, metallicity,
and luminosity function combinations of TRILEGAL at this Galactic position,
but the crucial feature --- the further restriction of the intrinsic NIR-MIR color distribution spreads --- is
clearly visible.  

\section{Application of the RJCE Method} \label{sec:dered}
\subsection{The RJCE Method Applied To 2MASS+{\it Spitzer}-IRAC Data} \label{sec:rjce}

One of the most extensive databases of NIR-MIR photometry comes
from combining 2MASS with the various GLIMPSE surveys of the Galactic midplane (see references in Churchwell et al.\ 2009). 
We will depend on the data in the GLIMPSE-I and Vela-Carina surveys for the remainder of this paper;
additional 
large sky surveys with the {\it Spitzer}-IRAC instrument afford the opportunity 
to apply the RJCE method to a large fraction of the dustiest directions in the Milky Way, but
this we leave to future work.

\begin{figure*} 
\begin{center}
\includegraphics[angle=90,width=1.0\textwidth]{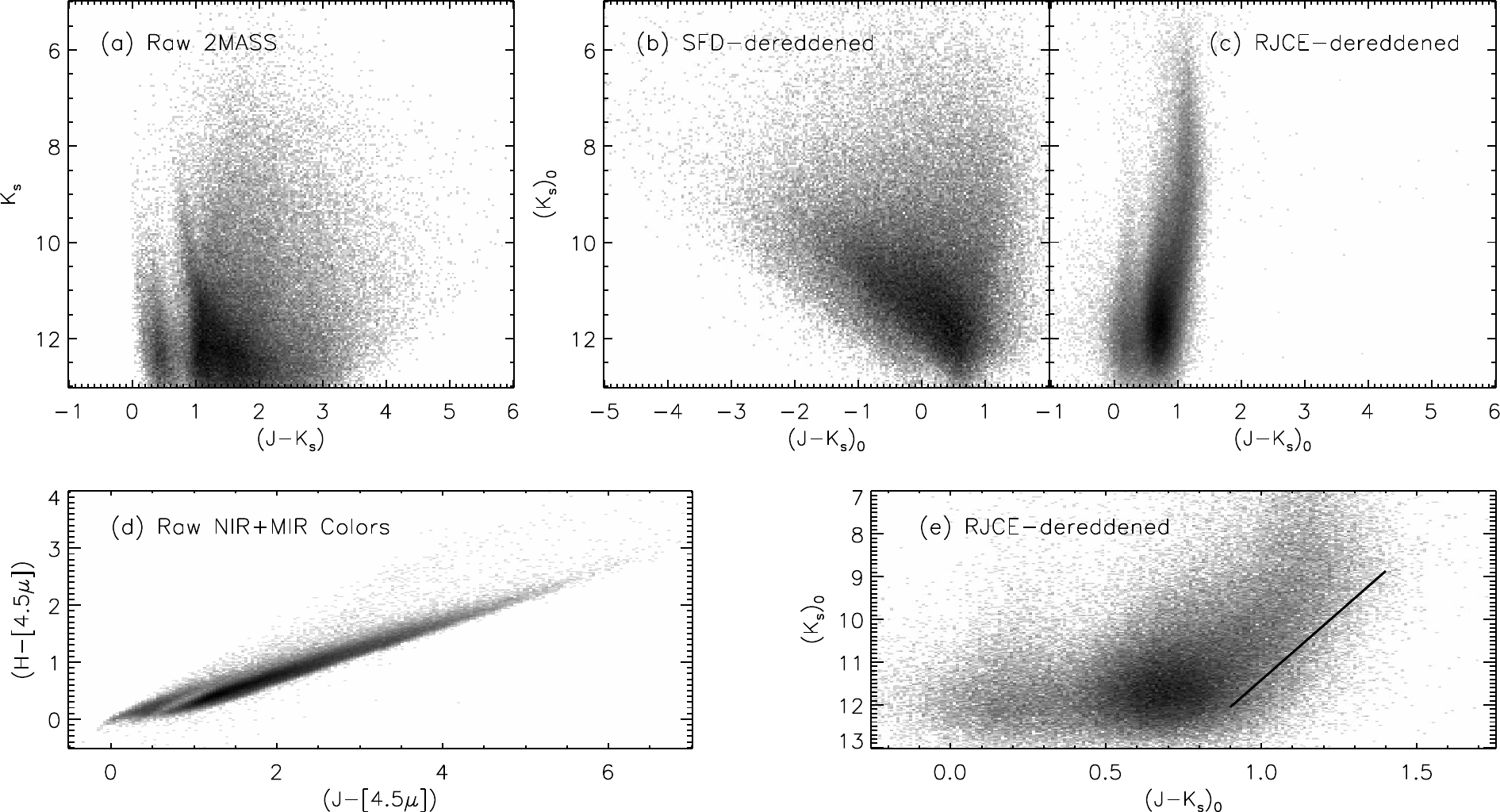}
\end{center}
\caption{(a) Raw 2MASS CMD for sources detected in a 4 deg$^2$ GLIMPSE field centered at $(l,b)=(307,0)^{\circ}$.  
(b) The same as (a) but corrected for reddening using the Schlegel et al.\ (1998, ``SFD'') reddening maps, 
where we have adopted the proscription of Bonifacio et al.\ (2000) for the treatment
of regions with $E(B-V)$ $>$ 0.1; to apply the SFD maps to the 2MASS CMD, we have assumed $R_V$ $=$ 3.1, 
$A(V)$/$A(K_s)$ $=$ 8.8 (Cardelli, Clayton, \& Mathis 1989), and $E(J-K_s)$/$A(K_s)$ $=$ 1.5 (Indebetouw et al.\ 2005).
(c) Same as (a) using the simplest RJCE dereddening algorithm described in \S\ref{sec:howitworks} ---
i.e. that all stars have the same $(H-[4.5\mu])_0$ color (adopted as 0.08).
The top three panels all have same color and magnitude spans.  
In contrast to panel (b), the three primary 2MASS field CMD fingers (main sequence
turn off at $[J-K_s]_0 \sim 0.2$, red clump at $[J-K_s]_0 \sim 0.7$, and
red giant branch at $[J-K_s]_0 \gtrsim 0.9$) are visible and restored to their proper colors 
by use of the RJCE method. 
(d) A 2MASS+{\it Spitzer} two-color diagram of the same field, 
showing the huge variation in reddening across 
this field and along the line of sight, which ranges from 0 to 6 magnitudes in ($J-[4.5\mu])$ (equivalent
to total extinctions spanning to roughly $A_V$ $\sim$ $30$). 
(e) Blow up of panel (c) to better show the primary CMD sequences.  
The line is the isochrone for a solar metallicity RGB from Ivanov \& Borissova (2002), 
shifted to 16 kpc.}
\label{fig:sfd_cmd}
\end{figure*}

Figure~\ref{fig:sfd_cmd} demonstrates the application of the simple, single color version of the RJCE method
via estimation of $E(H-[4.5\mu])$, after assuming that the intrinsic SEDs of
all stars have $(H-[4.5\mu])_0$ = 0.08;
for now we adopt this color --- typical for the evolved F, G and K stars that dominate at GLIMPSE
magnitudes
(Figure~\ref{fig:2cd}) ---
as the baseline 
$(H-[4.5\mu])_0$ color index for most stars.
To demonstrate how effectively the method works,
we have selected an extremely reddened demonstration field from the GLIMPSE
program (at $[l,b]$$=$$[307,0]^{\circ})$ that has a mean and maximum $E(B-V)$
given by SFD as 7.7 and 13.0 mag, respectively.
More significantly, as shown by the distribution of stellar colors 
in Figure~\ref{fig:sfd_cmd}d, this field has extremely variable {\it differential} infrared
reddening with 0 $\lesssim$ $E(J-[4.5\mu])$ $\lesssim$ 7 mag; this translates to a range of total 
extinctions spanning roughly 0 $\lesssim$ $A_V$ $\lesssim$ 30 mag. 

Figure~\ref{fig:sfd_cmd}a shows the raw 2MASS $([J-K_s],K_s)$ CMD for this demonstration field.
Two of the three nominally vertical ``fingers'' one sees in a typical 2MASS CMD of 
Galactic field stars --- one due to main sequence turnoff (MSTO) stars and the other due to
red clump (RC) stars
 --- are discernible in Figure~\ref{fig:sfd_cmd}a as sequences 
arcing to redder colors with increasing magnitude as both distance and the amount of cumulative foreground
dust increase.
At the faintest magnitudes shown, the red clump sequence broadens considerably due
to highly variable, patchy reddening.
The third stellar population sequence typically seen in the 2MASS CMD, that of more 
distant red giant branch (RGB) stars,
is completely smeared out in Figure~\ref{fig:sfd_cmd}a, because these stars typically suffer the most
total and differential reddening along the line of sight.

In general, the effectiveness of any
dereddening attempt can be gauged by the resulting color-magnitude coherence of typically
well-defined stellar population features like the RC, as well as the dereddened 
colors of the nominal stellar sequences. 
Figure~\ref{fig:sfd_cmd}b shows what happens to the
raw CMD when it is dereddened using the SFD extinction maps: The CMD actually
looks worse, in that (1) the primary CMD loci are
greatly smeared out, even more so than in the uncorrected CMD, and 
(2) the bulk of the stars have had their reddening grossly
overcorrected, as evidenced by the extraordinarily blue and obviously incorrect
colors for a large fraction of the stars.  In addition, (3)
the overall spread in colors in the SFD-corrected CMD is significantly larger than
for the uncorrected CMD.  That the SFD extinction maps tend to lead to overcorrections of colors 
in heavily extinguished Galactic midplane fields has been noted previously by
a number of studies (see discussion in \S\ref{sec:rjce-vs-sfd}).  These studies 
typically have found that the SFD maps were off by a simple scale factor; however, 
were this the only problem with the SFD maps, the coherence of the 
discernible CMD fingers in the raw CMD --- like that of the RC --- 
should be maintained when corrected 
for reddening by these maps, even if these sequences were shifted in color 
and orientation in the CMD.  The significantly smeared-out CMDs that result from
dereddening with the SFD maps show that these maps are not just off by a 
scale factor, but problematic in more nettlesome ways.  We will show the
extent of these issues with the SFD maps in highly extinguished 
regions in more detail in \S\ref{sec:rjce-vs-sfd}.

In contrast to the broadening of the distribution of stars in the SFD-corrected
CMD in Figure~\ref{fig:sfd_cmd}b, dereddening with the simple RJCE algorithm
described above --- using the Indebetouw et al.\ (2005) reddening law and
assuming
$(H-[4.5\mu])_0$ $=$ 0.08 for all stars ---
yields far superior results (Figure~\ref{fig:sfd_cmd}c), as evidenced by a much narrower overall
distribution of stars, with coherent MSTO, RC, and RGB stellar sequences  
corrected to their proper colors.  For example, the red clump is centered
at a color of $(J-K_s)_0 \sim 0.7$ mag, appropriate to disk metallicity RC
stars (e.g., Salaris \& Girardi 2002).  
Figure~\ref{fig:sfd_cmd}e is a close-up of the restored CMD to show more
clearly the RJCE-dereddened
colors of the distinct stellar population sequences. 
Figures~\ref{fig:sfd_cmd}c and \ref{fig:sfd_cmd}e demonstrate that the
primary stellar population sequences are not
only clearly visible and restored to their proper colors, but in the
case of the MSTO and RC they are each appropriately vertical and significantly
narrower (as seen in Figure~\ref{fig:sfd_cmd}c).  

The line drawn in Figure~\ref{fig:sfd_cmd}e is the isochrone for a solar metallicity 
RGB (from Ivanov \& Borissova 2002),
shifted to a distance of 16 kpc.  This line skirts
below the concentration of RC and RGB stars in this field and suggests not only that
a solar metallicity RGB is a reasonable match to the metallicity of these evolved
disk stars, but also that a true, sudden decrease in stellar density occurs near this distance
at about this longitude (that is, the sudden drop in the number of 
stars near the line is not due to sample incompleteness, 
but rather to a real drop in stellar density at the edge of the
MW disk for this line of sight; see discussion in \S\ref{sec:3dstars}).

\subsection{Estimation of RJCE Extinction Uncertainties} \label{sec:uncertainties}

The sources of uncertainty in extinction values derived with RJCE are 
(1) intrinsic scatter in the stellar color(s) used
to estimate reddening, (2) uncertainty in the stellar photometric measurements, 
and (3) variations and uncertainties in the extinction law used for reddening/extinction conversion.  
The first of these is assessed from the suite of Padova isochrones spanning the full available 
range of metallicity and age (i.e., those used in Figures~\ref{fig:2cd} and \ref{fig:2cd-tri}), 
yielding a maximum $\sigma$(H-[4.5$\mu$])$_0$ $\sim$ 0.04.\footnote{
 We emphasize that along a given line of sight, the stars observed
 will not span this full range of stellar parameters, so this $\sigma$ is more
 than a worst case scenario (e.g., compare Figures~\ref{fig:colhist} and \ref{fig:colhist-tri}).}
The photometric uncertainties of stars with both 2MASS $H$ and IRAC [4.5$\mu$] detections typically range from 0.02--0.12 mag in $H$
and 0.02--0.21 mag in [4.5$\mu$], with an average uncertainty of $\sigma_H$ $\sim$ 0.05 
and $\sigma_{[4.5\mu]}$ $\sim$ 0.1,
over the entire range of available magnitudes.
(Unlike the [5.8$\mu$] and [8.0$\mu$] IRAC bands, [4.5$\mu$] band photometry is largely unaffected by dust continuum emission;
Churchwell et al.\ 2009). 
We find these uncertainties to be uncorrelated with reddening, 
and these ranges describe even the more heavily reddened and crowded inner Galaxy.

Infrared extinction law variations, such as those assessed in Paper II and Indebetouw et al.\ (2005), 
may cause small systematic under- or over-estimates of the extinction
in particularly dense or diffuse ISM, respectively; 
however, the variation in $A(K_s)$/$E(H-[4.5\mu])$ measured in Paper II produces a $\lesssim$7\% 
possible error in the final $A(K_s)$ values (i.e., $\pm\sim$3.5\%).
We note that this does not include a possible systematic error due to the assumption of a constant $A_H$/$A_{[4.5\mu]}$,
but work targeting the non-bulge NIR extinction law (e.g., Stead \& Hoare 2009) has shown this variation to be small
compared to those at longer wavelengths.  
For very dense interstellar clouds (explicitly excluded from Paper II but studied in, e.g., Chapman et al.\ 2009),
which have the ``flattest'' (i.e., grayest) MIR extinction curves, the reddening-to-extinction ratio is only 7\% 
greater than the value we have adopted here.  And as we discuss in \S\ref{sec:limits}, these densest regions tend to 
suffer from far more significant problems that overwhelm this potential systematic error.

In summary, by combining the above sources of uncertainty, 
we estimate an approximate RJCE extinction uncertainty (in the $K_s$ band) of 
$\lesssim$0.11 mag for a typical {\it individual} star.
This {\it total} RJCE uncertainty is comparable to the contribution of
the intrinsic spread in the $(J-H)$ and $(H-K_s)$ 
colors alone in NIR-only color excess studies such as NICE.
And as we show below (and has been described in, e.g., Dobashi et al.\ 2008), 
the application of ``$N^{th}$ percentile'' extinction mapping
to large sets of stars significantly reduces the uncertainty of 
total extinction maps and rejects influence by statistical or genuine outliers.

\subsection{Comparison of Extinction Maps from the NICE, NICER, and RJCE Methods} \label{sec:cp-maps}

As discussed in \S\ref{sec:nice-r}, the development of NICE, NICER, and related methods has yielded variously effective
means for taking advantage of large area photometric surveys 
to make extinction maps in a variety of Galactic contexts.
These maps are limited in their
resolution by the density of  available stars and the level of smoothing over
stars --- the more stars averaged over, the lower are the influence of random errors 
from, e.g., limitations in the precision of the photometry, as well as
the innate dispersion and variation in intrinsic stellar colors.  
In the limit of large numbers of stars averaged over, these color excess methods are effective tools
for creating relatively reliable 
two-dimensional extinction maps.
The RJCE method is intended to improve on these
other methods by making more accurate estimates of reddening for {\it each}
star; this enables one to make extinction maps with arbitrarily small resolution, limited
only by the areal density of stars.  Nevertheless, it is useful and important to check the results 
of the various methods at a fixed resolution, one appropriate for all surveys.

\begin{figure} 
\includegraphics[width=0.48\textwidth]{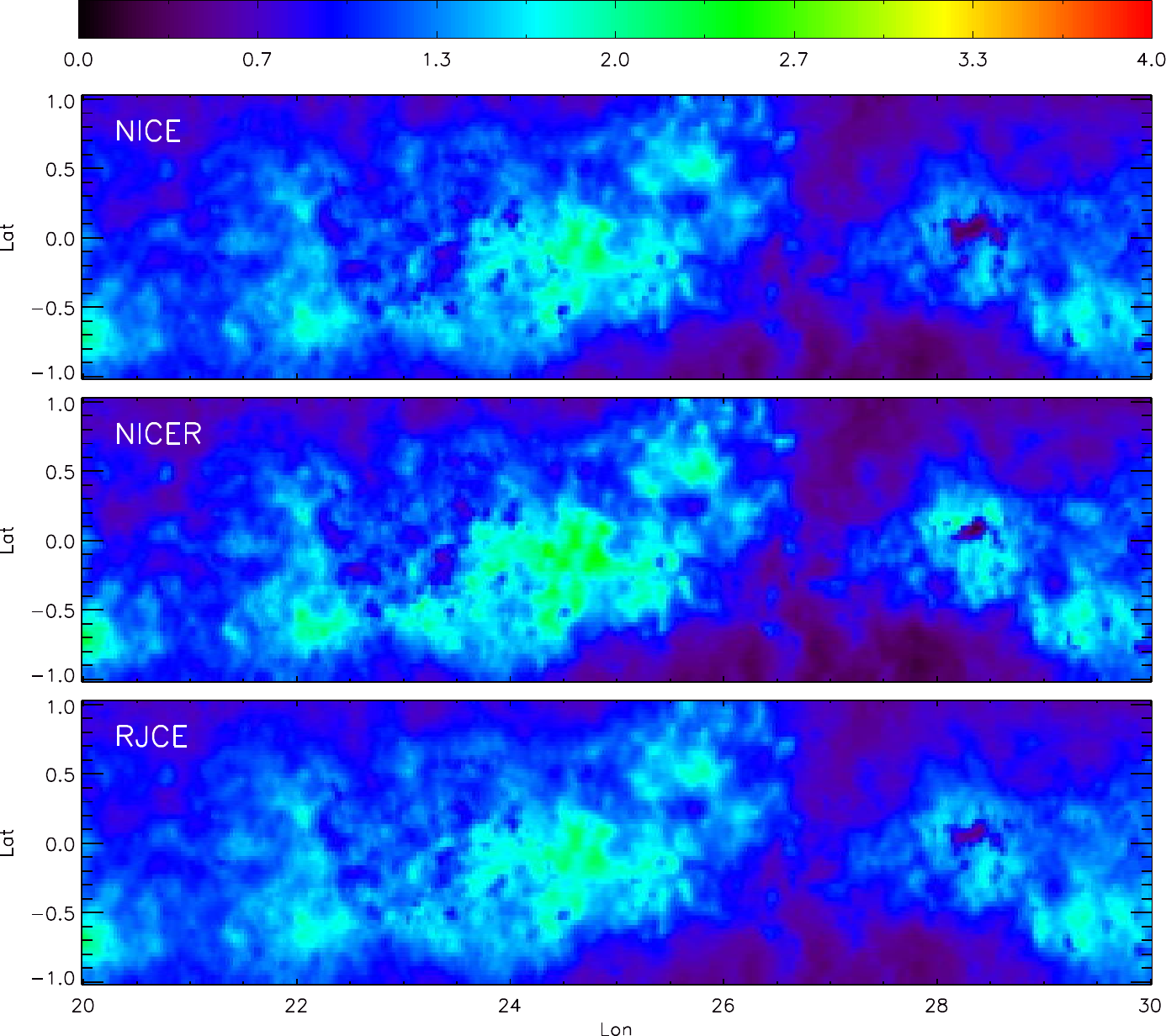}
\caption{$A(K_S)$ maps (where the
color indicates magnitudes of extinction according to the color
bar at the top) for $20^{\circ} \leq l \leq 30^{\circ}$ in the midplane
made using (a) the NICE method applied to $(H-K_s)$ colors, (b) the NICER method, using
a combination of $(J-H)$ and $(H-K_s)$ colors, and (c) the RJCE method, using $(H-[4.5\mu])$
colors.  In each case we use the median reddening of all stars in each pixel to derive the extinction in that pixel.}
\label{fig:cp-maps}
\end{figure}

Figure~\ref{fig:cp-maps} compares extinction maps of a segment of the GLIMPSE survey region
made with the NICE (using $[H-K_s]$ colors), NICER (using $[J-H]$ and $[H-K_s]$ colors), and RJCE methods,
with the latter making use of $(H-[4.5\mu])$ colors.  
For each map, we use all stars with uncertainties less than 0.5 mag 
in the set of photometric bands required for each method (e.g.,
$\sigma_H$ and $\sigma_{Ks}$ $\leq$ 0.5 for NICE).
In all cases we plot for each $2' \times 2'$
cell the median $A(K_s)$ extinction derived from color excess of all stars in the cell.
We adopt the following extinction relationships for the NIR-only maps, based on the Indebetouw et al.\ (2005)
extinction law and the mean stellar colors for giant stars 
(the dominant stellar type in the disk in surveys such as 2MASS and GLIMPSE) 
from the Girardi et al.\ (2002) isochrones:

\begin{equation}
\begin{array}{cc}
\displaystyle A(K_s) = 1.05 (J-H-0.76) & {\rm NICER} \\
\displaystyle A(K_s) = 1.82 (H-K_s - 0.13) & {\rm NICE(R)} \\
\end{array}
\end{equation}

\begin{figure}[!ht]
\begin{center}
\includegraphics[width=0.4\textwidth]{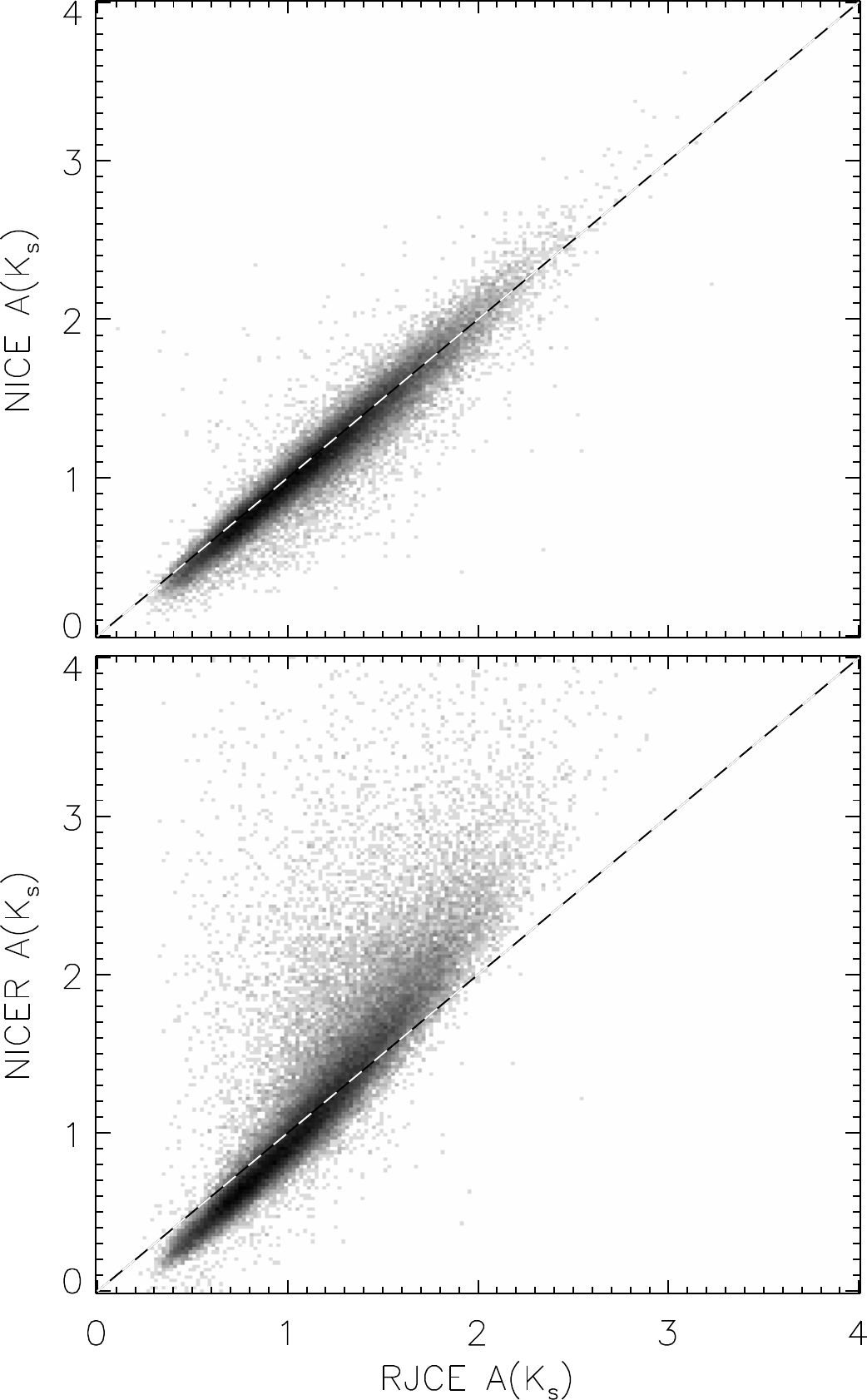}
\end{center}
\caption{A pixel-by-pixel comparison of the $A(K_s)$ extinction maps shown in Figure~\ref{fig:cp-maps}, derived from the
NICE and NICER methods (using NIR-only colors), to that derived from 
the RJCE method using $(H-[4.5\mu])$ colors.}
\label{fig:cp-ak}
\end{figure}

Visually, the maps created by the various color-excess methods strongly resemble each other.
To assess better their correspondence, we show in Figure~\ref{fig:cp-ak} the pixel values of 
$A(K_s)$ calculated from the NICE, NICER, and RJCE methods (Figure~\ref{fig:cp-maps}) 
plotted against one another.  A strong correspondence is found at most
$A(K_s)$ levels, although there is a trend at higher extinction for the NICER (and to a lesser extent, the NICE) $A(K_s)$ values
to be larger than those from the RJCE map.  
This result may be attributed to NICER's systematic overestimation of $A(K_s)$ towards red giant stars (explained more fully in \S\ref{sec:cp-cmds}), 
which would overestimate the measured extinction in each pixel by an amount proportional to the 
quantity of extinction.  A corresponding
underestimation of extinction towards foreground dwarf stars 
by NICE and especially NICER
would explain the difference in $A(K_s)$ pixel values 
at low extinction ($A[K_s]$ $\lesssim$ 0.8).

Another possible explanation for the non-unity correlation slope
in Figure~\ref{fig:cp-ak} is a problem with the adopted extinction law, and it is conceivable 
to get closer to a one-to-one correspondence if one were to adopt a modified extinction law
(e.g., if the equation $A(K_s) = 1.59 ([H-K_s] - 0.05)$ were used for the NICE-calculated
extinctions).  Note that it is not obvious {\it a priori} which total-to-selective extinction ratios
to modify to achieve this correspondence, and all three of the $E(J-H)$, $E(H-K_s)$ 
and $E(H-[4.5\mu])$ conversions to $A(K_s)$ could need modification depending on the 
actual extinction law.  
In principle, one could 
use distributions like that shown in Figure~\ref{fig:cp-ak} to rederive 
the {\it relative} color extinction ratios for any particular
line of sight; this is the essence of the work presented
in Paper II.
It should be
noted that without consistency in these reddening-to-extinction conversions, color-excess 
dereddening methods relying on multiple colors, such as NICER, may have additional 
systematic errors introduced.

\begin{figure}[!h]
\begin{center}
\includegraphics[angle=0,width=0.4\textwidth]{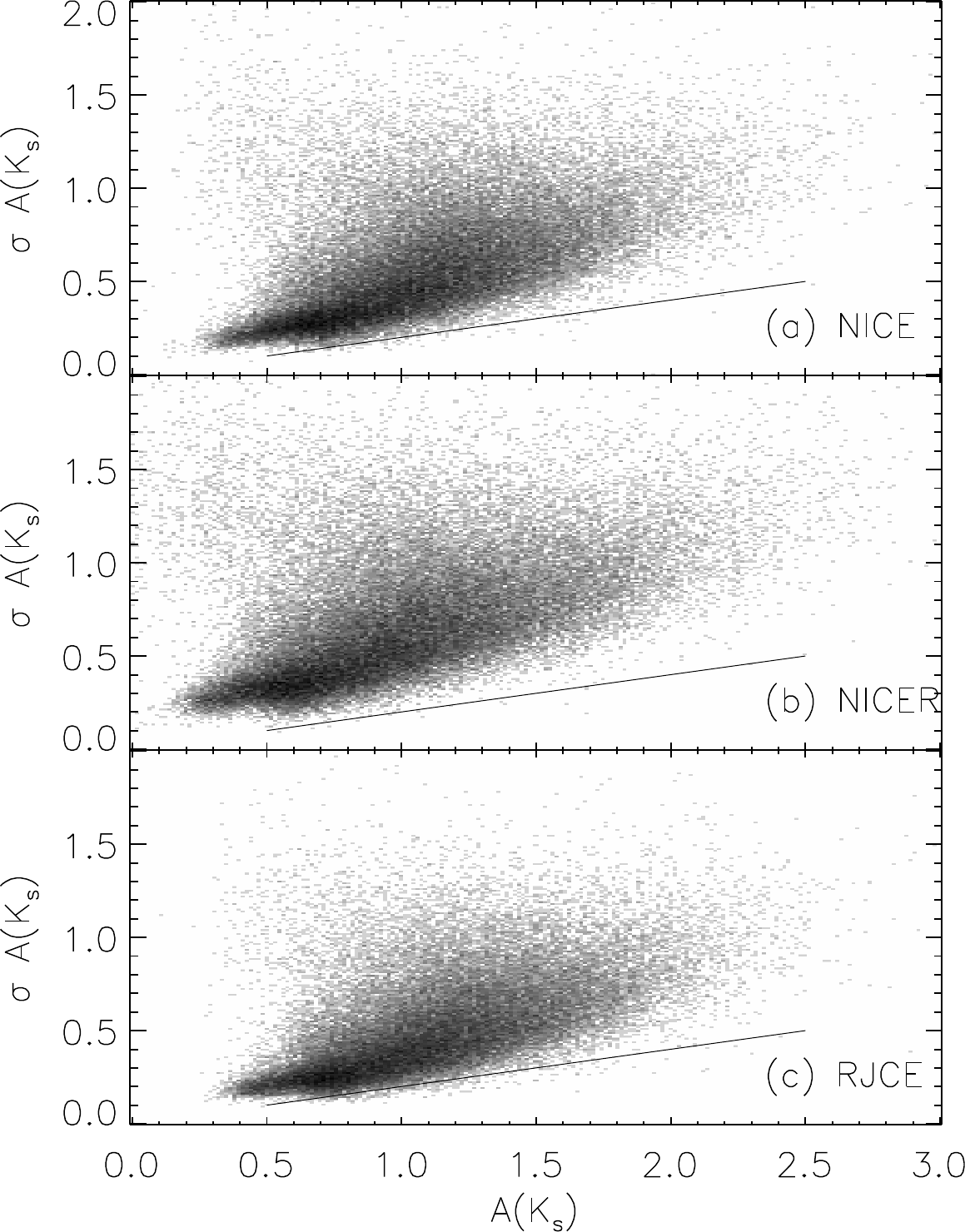}
\end{center}
\caption{The dispersion in the $K_s$ extinction, $\sigma(A_{Ks})$, as a function of
$A(K_s)$, for the pixels in the maps shown in Figure~\ref{fig:cp-maps}.  Panel (a) shows the results for the NICE map in 
Figure~\ref{fig:cp-maps}a, panel (b) shows the results for the NICER map in Figure~\ref{fig:cp-maps}b, and panel (c)
shows the results for the RJCE map in Figure~\ref{fig:cp-maps}c.  The solid line is identical in 
each panel and is included to aid comparison.}
\label{fig:disp}
\end{figure}

The relative reliability of color excess maps is commonly assessed by
plotting the dispersion in extinction values for stars within a map pixel as a function
of the derived extinction in that pixel
(e.g., Lada et al.\ 1994; Lombardi et al.\ 2008).
In part, this type of analysis has been used to ascertain the degree to which there is
structure in the dust distribution on angular scales smaller than the pixel size. 
We note that previous authors have used the visual extinction, $A(V)$ as the reference, though
$A(K_s)$ is just as useful for this purpose, and more naturally and reliably determined from 
infrared colors.  In Figure~\ref{fig:disp} we compare the dispersion in $A(K_s)$ as a function of
$A(K_s)$ itself for the maps shown in Figure~\ref{fig:cp-maps}.  The solid line is an approximate
match to the bottom envelope of the RJCE relationship and is included in all three plots to guide the eye in comparison. 

Two features of this comparison are evident: (1) the slope of the RJCE dispersion plot
is lower than that of the NICE and NICER dispersion plots, 
and (2) the {\it spread} in the dispersion
at each $A(K_s)$ is lower in the RJCE case than in the others.  Both features can
be attributed to the greater fidelity of the derived color excesses in the case of RJCE, 
due to the lower dispersion in intrinsic $(H-[4.5\mu])_0$ stellar colors compared to that for intrinsic NIR 
colors (see Figures~\ref{fig:2cd} and \ref{fig:2cd-tri}).    
However, all of the dispersions are much greater than the expected extinction uncertainty (\S\ref{sec:uncertainties}),
which suggests that at this map resolution, the dispersion is dominated by sub-pixel-scale extinction variations
(see \S\ref{sec:3dmap} and Figure~\ref{fig:pop-disp} for an example of how RJCE can greatly ameliorate this problem
by restricting the spatial extent of the stars tracing the extinction).

\subsection{Comparison of CMDs from the NICE, NICER, and RJCE Methods} \label{sec:cp-cmds}

As described in \S\ref{sec:intro}, until now, color excess techniques have been primarily focused
on creating extinction maps,
whereas we are also interested in color-magnitude diagrams.  These can be used not only to interpret
the stellar populations along a line of sight, but also to improve our detailed 
understanding of the distribution of dust by placing 
reddened and dust-extinguished stars at specific and accurate photometric parallax distances.
In this context, the NICE/NICER family of methods 
are less useful approaches than the RJCE method because of
their less accurate assumptions about intrinsic stellar colors, their averaging of reddenings
over many stars, 
and their inclusion of photometric bands --- particularly
the NIR $J$-band or bluer bands --- 
that are highly sensitive not only to reddening by dust but also the intrinsic SEDs of the stars themselves.

\begin{figure} 
\begin{center}
\includegraphics[angle=90,width=0.45\textwidth]{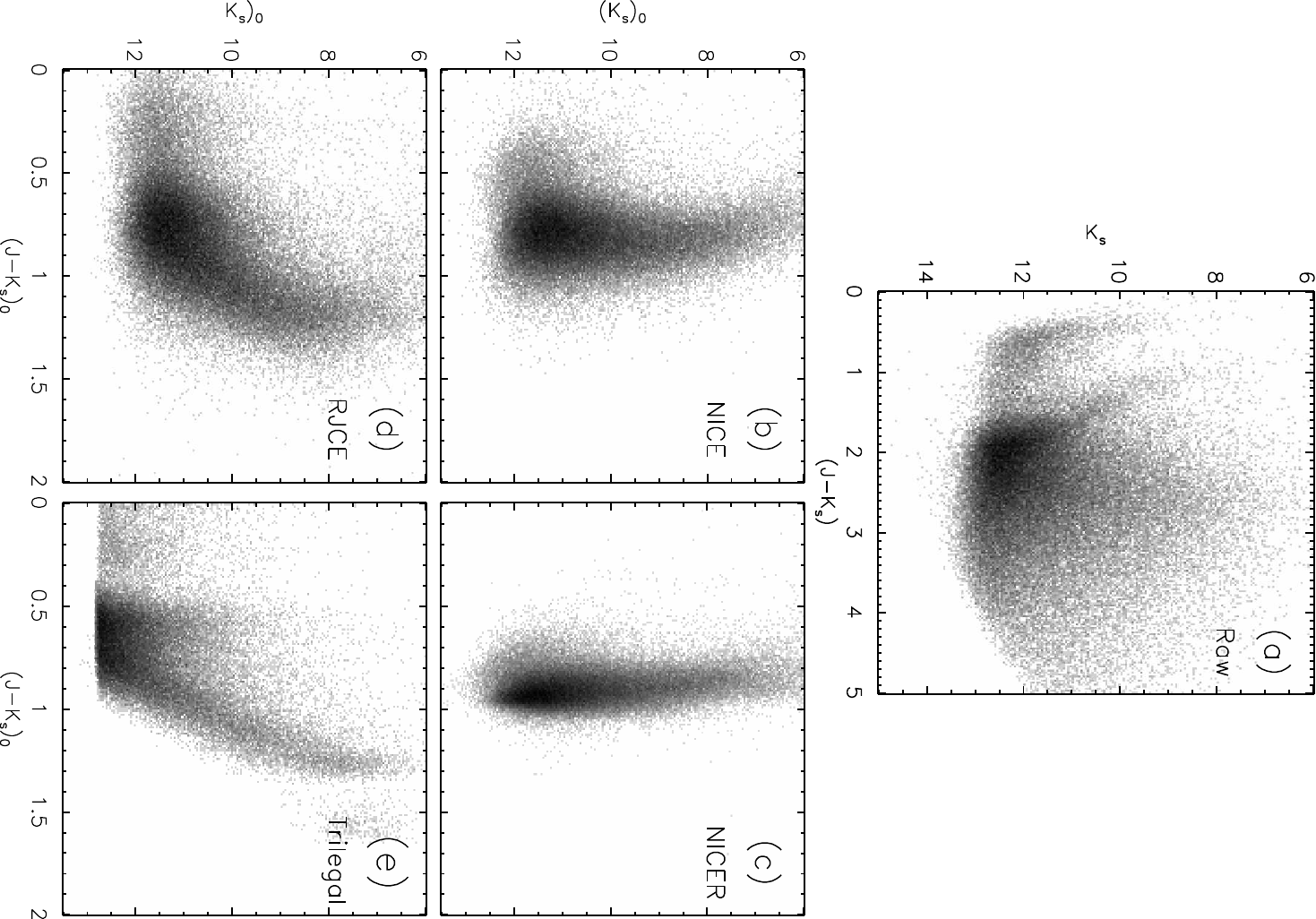}
\end{center}
\caption{
Color magnitude diagram for a 4 deg$^2$ field centered at $(l,b)$ = (42,0)$^\circ$ with data from the 2MASS and GLIMPSE catalogs.
Panel (a) shows the raw photometry with no quality cuts.
In panels (b) and (c), the NICER method has been applied using the equations reported in Lombardi \& 
Alves (2001), assuming $(J-H)_0= 0.76$ and $(H-K_s)_0 = 0.13$
(from isochrones of RC and RGB stars with [Fe/H] $\gtrsim$-1; Girardi et al.\ 2002)
 for all stars.  
The assumed colors for NICER 
are probably different than the (unreported) ones 
used by Lombardi \& Alves (2001), but assuming different colors only slightly changes the 
position of the final dereddened CMD's loci, not their overall {\it shape}.  
In panel (d), the RJCE method has been applied to deredden the CMD, assuming $(H-[4.5\mu])_0= 0.08$
(also adopted from the Padova isochrones)
for all stars.  For the dereddened CMDs, the extinction law of Indebetouw et al.\ 2005 
was used to convert reddenings to $A(K_s)$, and only stars with a detection in all seven NIR-MIR bands are shown.
Panel (e) contains the TRILEGAL model CMD (Girardi et al.\ 2005). for the same region of the sky coverage,
where we have assumed all default parameters in Version 1.4 of that model
except for 
extinction (adopted as {\it A(V)} = 0 for all stars) and with a limiting magnitude of $K_s=12.5$.}
\label{fig:cmds}
\end{figure}

These limitations are
demonstrated in Figure~\ref{fig:cmds}, which compare CMDs dereddened by
these various infrared excess 
techniques.  Figure~\ref{fig:cmds}a shows a raw 2MASS, ($[J-K_s], K_s$) CMD for a field
of area 4.0 deg$^2$ at $(l,b) = (42,0)^{\circ}$ and having strong and variable extinction, 
with 0 $<$ $A(K_s)$ $<$ 4.  As in Figure~\ref{fig:sfd_cmd}a, the CMD shows arcing trends toward redder
colors at fainter magnitudes as a result of increased dust reddening along the line of sight.
Figure~\ref{fig:cmds}e shows the expected CMD for the same field without reddening effects, as
predicted by the TRILEGAL stellar populations model\footnote{http://stev.oapd.inaf.it/cgi-bin/trilegal} (version 1.4, Girardi et al.\ 2005).  
We have added photometric errors in accordance with those expected from the $m-\sigma_m$ relationship of the 2MASS data. 
(Note that these errors have been applied to the 
{\it intrinsic} stellar colors and magnitudes; thus the amount of induced scatter in the colors
is slightly less than would be
the case for stars with fainter, extinguished magnitudes.)
The TRILEGAL model shows the three principal stellar population loci
discussed in \S\ref{sec:rjce} and seen in Figures~\ref{fig:sfd_cmd}c and \ref{fig:sfd_cmd}e: 
main sequence, red clump and red giant stars (from blue to red, respectively).  
The point of this model comparison
is to examine the variously dereddened 2MASS CMDs for the correctness of the 
color distributions of the primary stellar populations
and to see how the techniques affect the overall CMD appearance.

In Figure~\ref{fig:cmds}b we show a $([J-K_s]_0,[K_s]_0)$ CMD derived using the NICE method, applied to the observed
$(H-K_s)$ colors of each star.  As in the studies described in \S\ref{sec:nice-r}, 
we adopt a mean intrinsic color of $(H-K_s)_0 = 0.13$  for each star; this assumption
has the effect
of strongly compressing the derived $(J-K_s)_0$ colors.  In particular, the reddest, brightest giants
are ``over-dereddened'' because these stars are intrinsically redder than the adopted mean
color, and the NICE method attributes the observed redder colors for these stars {\it entirely}
to reddening.  
The opposite happens for the main sequence stars, which have intrinsic colors bluer
than $(H-K_s)_0 = 0.13$ and are therefore insufficiently dereddened by NICE.

This color compression is even more pronounced with a NICER dereddening; 
the latter technique uses both $(J-H)$ and $(H-K_s)$ colors to deredden, which results
in an even greater homogeneity in the derived $(J-K_s)_0$ colors (Figure~\ref{fig:cmds}c).
Thus, the three major stellar populations (MS, RC, RGB) have
been unrealistically compressed to essentially a single color (effectively representing the
SED corresponding to the adopted ``mean'' intrinsic colors), removing any chance of 
recovering a star's stellar type or luminosity class information.

Figure~\ref{fig:cmds}d applies the RJCE method using the $(H-[4.5\mu])$ color assumption (e.g., as in
Figure~\ref{fig:sfd_cmd}c and \ref{fig:sfd_cmd}e).  Of the various dereddening methods, the RJCE method, which makes no assumptions about
the stellar type in its application but uses only the homogeneity of $(H-[4.5\mu])_0$
colors for all stellar types, produces the most sensibly-dereddened NIR CMD, as 
seen by comparison of Figure~\ref{fig:cmds}d to Figure~\ref{fig:cmds}e.  In this panel, all of
the primary loci have the same shape and color distribution as the corresonding loci in 
the TRILEGAL zero-extinction CMD.

One advantage of a relatively cleanly dereddened CMD 
is that one can select, with relatively high purity, specific stellar types to serve as tracers
of the dust.  For example, one could select RGB stars --- the vast majority of stars redward
of $(J-K_s)_0$ $\sim$ 0.85  in the CMD of Figure~\ref{fig:cmds}d ---
as the most distant stars at any
particular magnitude, 
and thus those most useful for total extinction maps.
Or, one could focus on red clump stars (the dominant stellar type at $(J-K_s)_0 \sim 0.55$--$0.8$), 
which are not intrinsically as bright as RGB stars
but which make good standard candles with very little dependence of the intrinsic luminosity 
on metallicity (e.g., Salaris \& Girardi 2002). 
See an initial exploration of this method in \S\ref{sec:3dmap}.  

\subsection{Photometric Limitations in High Reddening Regions} \label{sec:limits}

The observant reader may have noticed suspiciously sharp ``holes'' in the 
color excess extinction maps (Figure~\ref{fig:cp-maps}) at the center of the highest extinction regions.
Such holes are artifacts created by magnitude limits in the adopted photometric catalog 
(as this section will demonstrate, the limitation is generally the 2MASS contribution to the GLIMPSE catalog, 
so that all three of the maps shown are affected similarly).
Obviously, if the dust in a particular cloud is dense enough, the total extinction 
can be sufficient to increase the magnitudes of any tracers behind the cloud
beyond the magnitude limit of the catalog.  In the event that no stars beyond
the cloud are visible, the dense region of the cloud cannot be detected by
any color excess method,
and that region of a two-dimensional map is 
either devoid of stars (if the cloud is close enough) or
dominated by the lower extinction measured along the line of sight to stars foreground to the cloud.  
The only remedies to this sort of problem are to (1) ignore the data given in such
areas (with these regions identified and flagged in some way), (2) obtain deeper photometry,
or (3) use photometric bands with a lower sensitivity to dust with which stars
beyond the cloud may be seen.

\begin{figure}[!h]
\begin{center}
\includegraphics[width=0.45\textwidth]{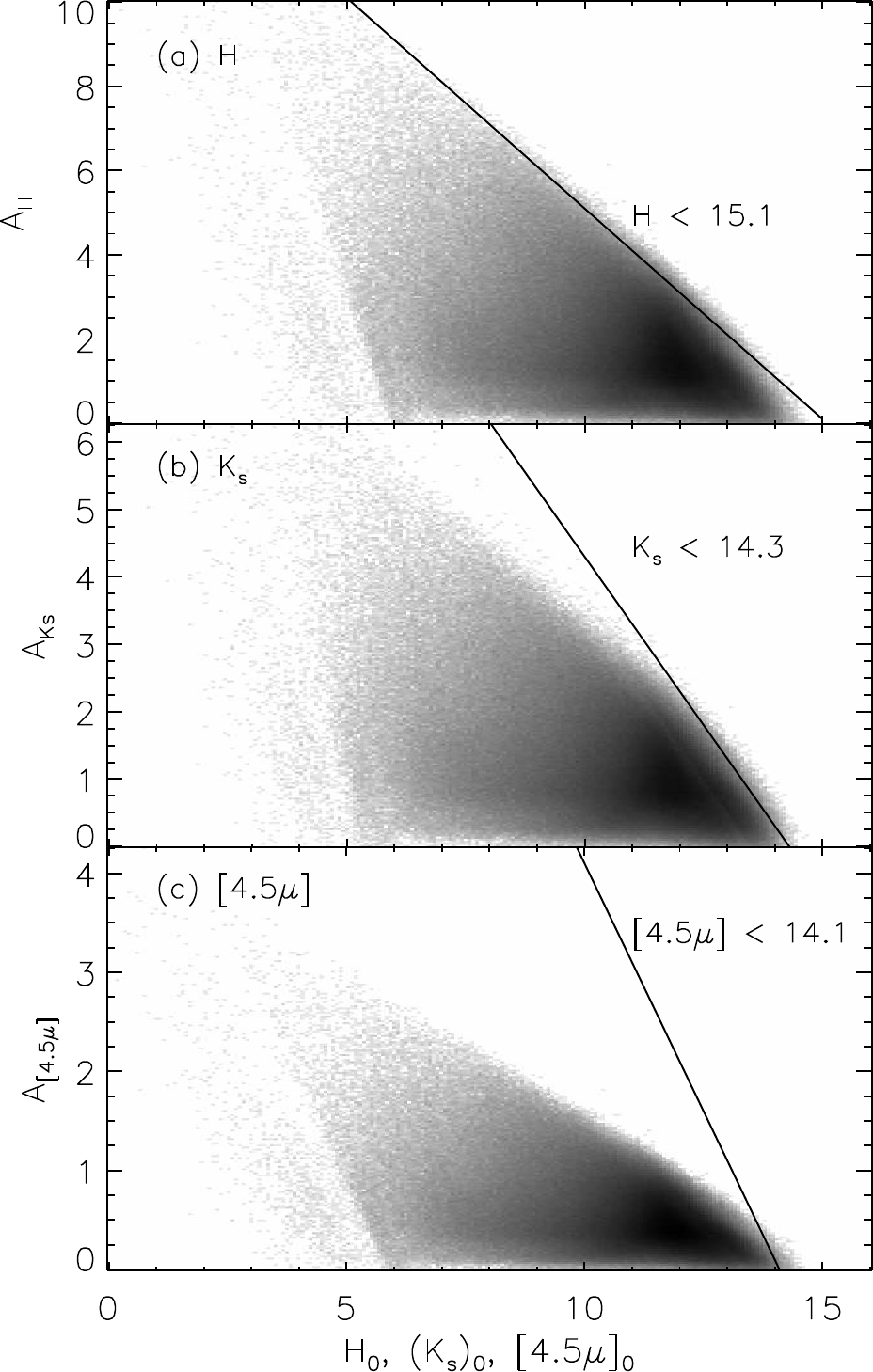}
\end{center}
\caption{The distribution of derived extinction to stars
as a function of their RJCE-dereddened magnitude in the (a) $H$-band, (b) $K_s$-band,
and (c) IRAC $[4.5\mu]$ band for the region shown in Figure~\ref{fig:cp-maps}.  
The diagonal line in each panel represents a hypothetical observed magnitude 
limit (indicated by the values shown in the panel) near which the faintest stars contributing
to the Figure~\ref{fig:cp-maps} maps would lie if the dominant limiting magnitude in the catalog
were that given by that particular filter.  As may be seen, the primary
limit on the input catalog to the Figure~\ref{fig:cp-maps} maps is the photometric precision of the $H$-band 
filter of 2MASS.}
\label{fig:limits}
\end{figure}

So far in this paper we have focused primarily on extinction maps and 
dereddening based on band-merged 2MASS+{\it Spitzer}-IRAC data.  
We have generally applied selection limits to this catalog based on choosing stars with photometric
uncertainties smaller than 0.5 mag in the bands required for each particular method being used.
This translates to a primary limitation being placed by the 2MASS data, as demonstrated
in Figure~\ref{fig:limits}.  There we present for those stellar data mapped in Figure~\ref{fig:cp-maps} 
the amount of extinction measured foreground to each star in each of the $H$, $K_s$, and [4.5$\mu$] bands
as a function of the RJCE-corrected 
$H$, $K_s$, and [4.5$\mu$] magnitudes of that star, respectively.  
The extinctions in each panel 
were actually measured via the color excess in $(H-[4.5\mu])$, but
translated from the selective to total extinction in the photometric band shown on the abscissa
using the Indebetouw et al.\ (2005) extinction law.
In Figure~\ref{fig:limits}, the catalog's {\it intrinsic} magnitude limit 
expected from the apparent magnitude limit of the filter explored in each panel 
is linear and shown by the
diagonal line with a slope of $-1$.
So, for example, it may be seen in Figure~\ref{fig:limits}a that no star with $H_0$ + $A(H)$ $\gtrsim$ 15 
is contributing to the maps in
Figure~\ref{fig:cp-maps}.  This appears to be the dominant
limitation in the Figure~\ref{fig:cp-maps} maps, because
Figures~\ref{fig:limits}b and \ref{fig:limits}c show that, for the most part,
stars with $(K_s)_0$ + $A(K_s)$ $\lesssim$ 14.3 or $[4.5\mu]_0$ + $A([4.5\mu])$ $\lesssim$ 14.1
would be detected were they not already excluded by the $H$-band magnitude limit.
However, closer inspection reveals that
it is actually the $K_s$ band photometry that seems to be the primary catalog 
limitation at low ($A[K_s]$ $\lesssim$ 2) levels.
Figure~\ref{fig:limits} demonstrates that color-excess maps made using 2MASS 
data, either entirely or partly,
will have trouble probing the highest extinction regions, with an extinction limit dependent on the 
distance of the tracer (indirectly, via its unextinguished apparent magnitude).  

\begin{figure} 
\begin{center}
\includegraphics[width=0.48\textwidth]{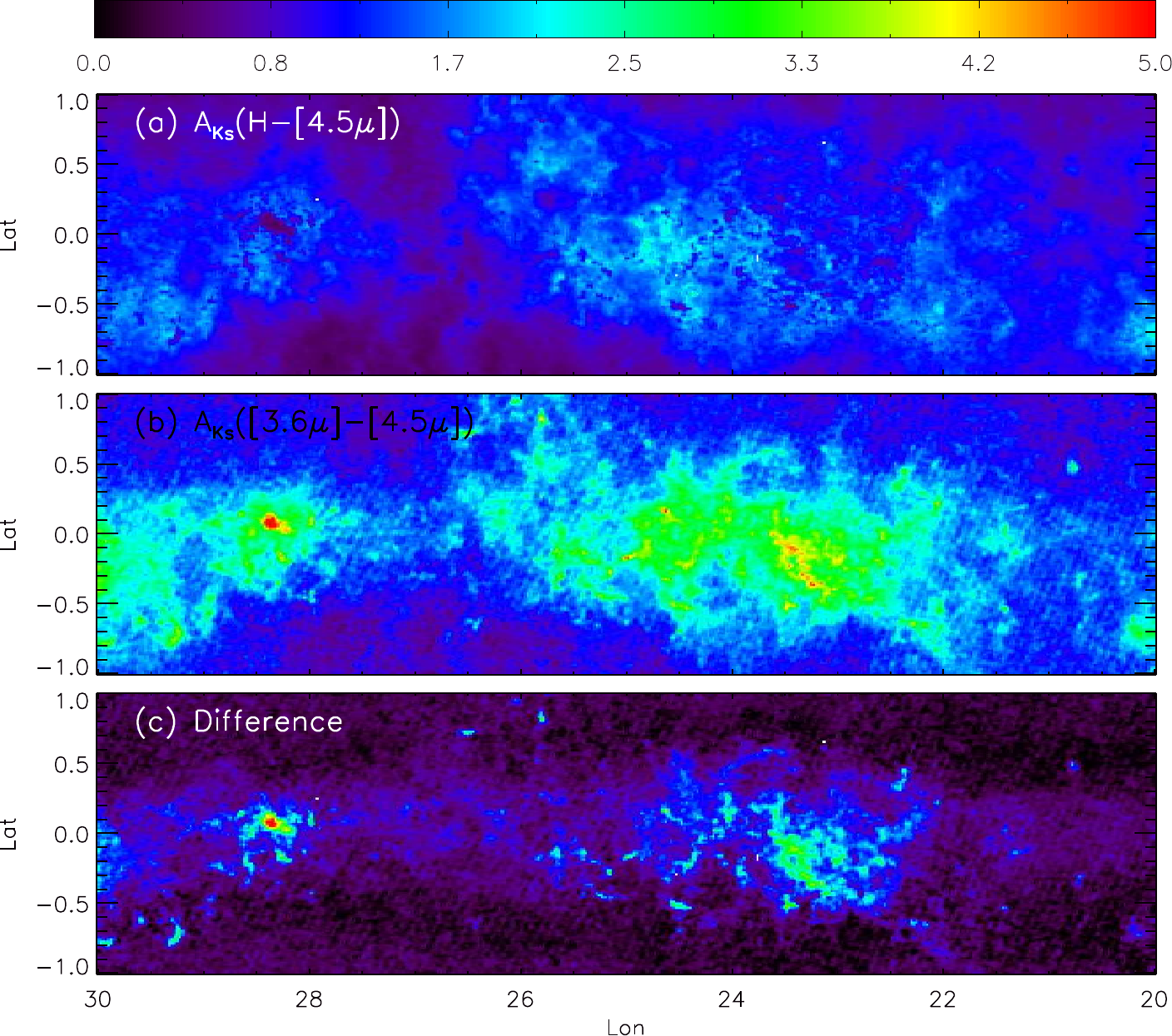}
\end{center}
\caption{Total extinction maps (in units of $A(K_s)$)
made by the RJCE method, with use of the (a) $(H-[4.5\mu])$
colors and (b) $([3.6\mu]-[4.5\mu])$ colors as the selective extinction measure (where we 
have adopted $([3.6\mu]-[4.5\mu])_0$ = -0.1 as the baseline Rayleigh-Jeans stellar color).  
In each case, the stars used 
are those satisfying a 0.5 mag
photometric uncertainty limit in only the two filters used 
in the color-excess measurement.  Panel (c) shows the difference between the 
maps in panels (a) and (b).  The few white pixels indicate bins containing $<$2 stars.  }
\label{fig:rjce-diff}
\end{figure}

On the other hand, as suggested by the severe difference between the potential and
actual limits of the [4.5$\mu$] data shown in Figure~\ref{fig:limits}c, use of IRAC data alone
holds great promise for probing these highly extinguished regions.  This is highlighted
vividly by the RJCE-generated maps shown in Figure~\ref{fig:rjce-diff} made using 2MASS and GLIMPSE
data for (a) $(H-[4.5\mu])$
colors and (b) $([3.6\mu]-[4.5\mu])$ colors.  Again, we have used catalogs 
restricted only by the photometric uncertainty limits of the relevant filters; the greater depth of the IRAC photometry,
as well as the lower total-to-selective extinction ratio at these wavelengths,
enables much greater sensitivity to dense clouds using only MIR filters.  As the difference map (Figure~\ref{fig:rjce-diff}c)
makes especially clear, the IRAC-only map is capable
of probing the dense cores of cloud complexes that appear only as artificial ``holes" in the 
$(H-[4.5\mu])$ extinction map due to
stars behind the dense clouds being extinguished out of the 2MASS catalog.
It is also obvious that greater overall extinction levels are detected in the more deeply probing ([3.6$\mu$]-[4.5$\mu$]) map.

While the example shown in Figure~\ref{fig:rjce-diff} might suggest exclusive use of MIR (e.g.,
IRAC) photometry as the ideal solution to creating extinction maps, there are several
mitigating factors that need to be considered: (1) As shown in Figures~\ref{fig:colhist} and \ref{fig:colhist-tri},
the intrinsic ([3.6$\mu$]-[4.5$\mu$])$_0$ color range of
stars is apparently wider than, say, that of longer 
wavelength baseline NIR-MIR colors, like $(H-[4.5\mu])_0$ and $(K_s-[4.5\mu])_0$.  Of 
course, this might be overcome by the use of longer wavelength combinations of
MIR filters.  However, the GLIMPSE 
photometric uncertainties at 5.8$\mu$m and 8$\mu$m (IRAC channels 3 and 4)
are greater than at 3.6$\mu$m and 4.5$\mu$m, 
the contribution from dust continuum emission is more significant,
and the extinction law ($A_\lambda$/$A_{Ks}$)
at these wavelengths is significantly more variable throughout the Galactic disk (Paper II).
(2) Because of the still relatively narrow
intrinsic color spread for stars in MIR colors, like $([3.6\mu]-[4.5\mu])$, 
there is virtually no stellar type information that can be gleaned for normal stars using
MIR colors {\it only}.  Thus, one loses almost all stellar population
information (i.e., there is no useful CMD in these filters) that one could use 
to prune one's tracer sample to specific tracers (e.g., the intrinsically
brightest, farthest stars at each magnitude).
(3) Being more weakly affected by dust means that the MIR has less sensitivity to
low levels of extinction, where it will generally be less reliable at deriving $A(K_s)$ for a given
precision in the photometry.
All of these suggest that hybrid schemes, whereby NIR+MIR (e.g, 2MASS+GLIMPSE)
photometry is used for low extinction regions and MIR photometry alone is used 
in higher extinction regions (where the NIR photometry is limited in reach)
might be the most optimal way to utilize color-excess extinction mapping when depending on
databases like 2MASS and GLIMPSE.  
Obviously, with the advent of deeper, publicly-available NIR surveys in the future (such as UKIDSS),
the limitations of the NIR photometry seen here can be mitigated; given the 
limited sky coverage for high-resolution MIR photometry, such surveys have the capability of improving
the extinction reach of NICE-like mapping, which can be used outside of the regions probed
by IRAC.  Of course, as discussed at length in \S\ref{sec:cp-cmds}, higher precision color-excess dereddening
will always be obtained by combining NIR and MIR photometry.
The recent and future releases of the WISE all-sky MIR photometry (Wright et al.\ 2010) will provide an opportunity to apply
reliable RJCE extinction mapping and CMD corrections across the entire sky, particularly in out-of-plane,
high-reddening regions not observed with {\it Spitzer}-IRAC.

\section{Comparison of RJCE Extinction Maps to
Those Using Non-Color-Excess Proxies for Dust Extinction} \label{sec:cp-othermethods}

\begin{figure*}[!htpb]
\begin{center}
\includegraphics[angle=0,width=1.0\textwidth]{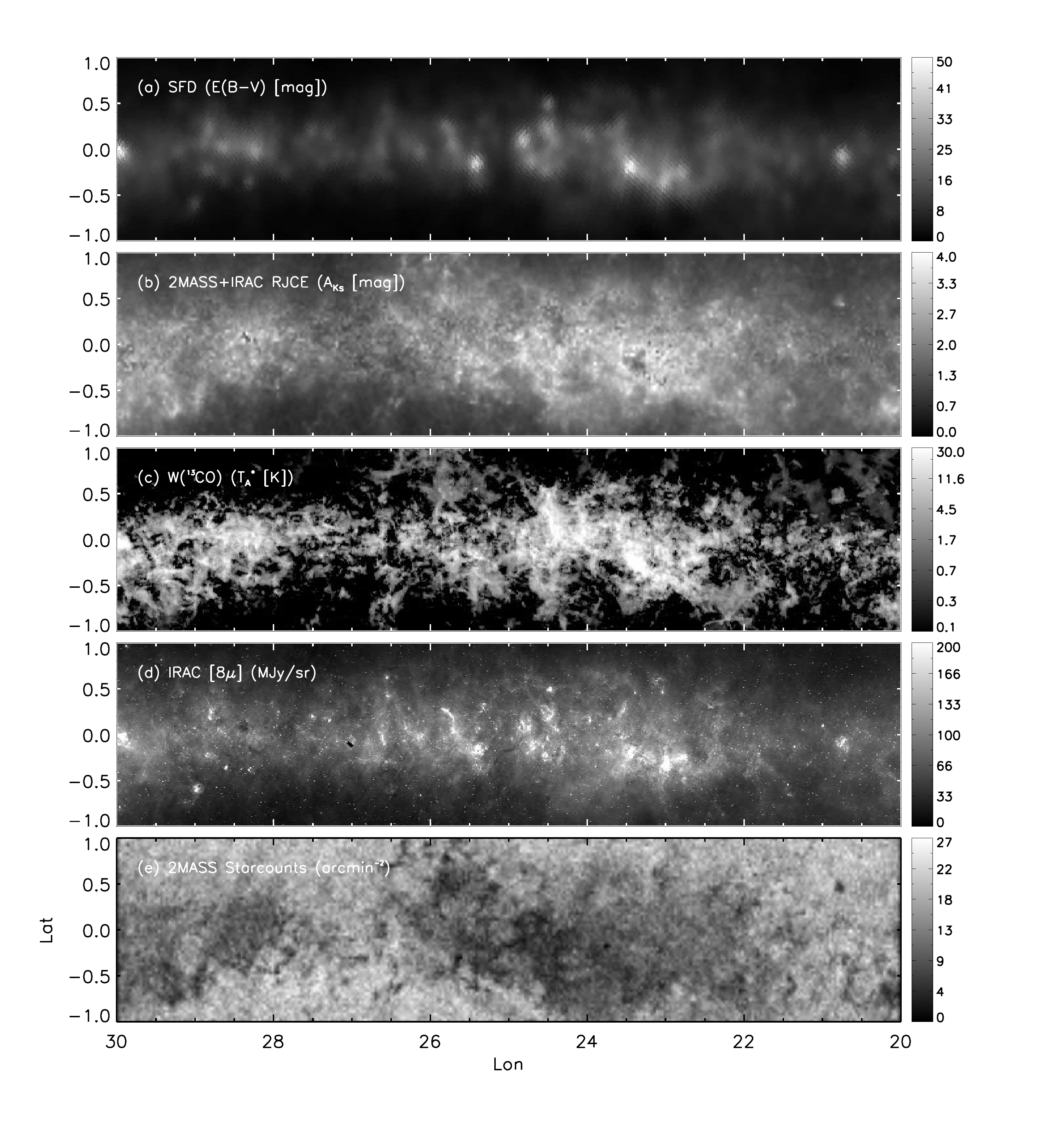}
\end{center}
\caption{
A portion of the Galactic plane contained within GLIMPSE, seen in five separate views.   
(a) The Schlegel, Finkbeiner, \& Davis (1998) reddening map, which has roughly 6 arcmin resolution.  As in Figure~\ref{fig:sfd_cmd},
pixels with $E(B-V)$ $\geq$ 0.1 
have been modified according to the prescription in Bonifacio et al.\ (2000; see text).  
The linear greyscale (shown to the right, with units of mags) is set to black for $E(B-V) = 0$ 
and to white for $E(B-V) = 50$ 
($A_V$ $\gtrsim$ 150).
(b) A map of the extinction using the RJCE method (based on $(H-[4.5\mu])$ colors) with $\sim$2 arcmin resolution, 
where each pixel displays the extinction for the 90th percentile strongest extinguished star within that pixel.
The linear greyscale is set to black for $A(K_s) = 0$ and to white for $A(K_s) = 4$.
As discussed in \S\ref{sec:limits}, for the very most extinguished regions there are no stars remaining in our
magnitude-limited sample, and at these positions the map has a ``hole" in it, which appears
black in this image.  Only a small number of pixels are so affected.
(c) A map of the $^{13}$CO ($J=1 \rightarrow 0$) integrated emission (all velocities) from the GRS survey (Jackson et al.\ 2006).
The linear greyscale is set to black for no signal and to white for an antenna temperature of 30 K.  
(d) An 8$\mu$m emission map from {\it Spitzer}-IRAC taken as part of the GLIMPSE-I survey (Benjamin et al.\ 2003).    
The linear greyscale is set to black for no emission and to white for a diffuse brightness of 200 MJy/sr.
Bright stars appear as small white dots in the image and are not included in the diffuse brightness scaling.
(e) The surface density of stars appearing in the bandmerged 2MASS+GLIMPSE catalog brighter than a magnitude limit of
$K_s = 14.3$.  
The linear greyscale is set to black for 2 stars arcmin$^{-2}$ and to
white for 37 stars arcmin$^{-2}$.}
\label{fig:cp-maps2}
\end{figure*}

In \S\ref{sec:cp-maps} we showed and compared extinction maps created by various IR color excess methods. 
These comparisons only give indications of the relative performance of 
these maps with respect to one another.  But are such maps reliable indicators of dust distributions
in the first place?  In this section we compare the color-excess-generated extinction maps
--- specifically those generated through the RJCE method --- with other ``proxy tracers" for dust extinction.  
From this comparison we conclude not only that the RJCE-generated maps are reliable indicators of the
two-dimensional distribution of dust down to rather fine scales, but also that the RJCE method currently
may well be the most reliable means to generate extinction maps on small and large scales.
The comparisons made in this section are summarized in Figure~\ref{fig:cp-maps2}, which shows a series
of maps of the GLIMPSE region from $20^{\circ} \le l \le 30^{\circ}$.
Figure~\ref{fig:cp-maps2}b is the RJCE extinction map, constructed from ($H-[4.5\mu]$) colors of RC and RGB stars, using
for each $2' \times 2'$ pixel the 90th-percentile-measured stellar extinction in that pixel. 

\subsection{Comparison to $^{13}$CO ($J=1\rightarrow 0$) Emission} \label{sec:rjce-vs-co}

Figure~\ref{fig:cp-maps2}c shows the integrated $^{13}$CO ($J$=0$\rightarrow$1) emission in this region of the Galactic midplane, with data taken
from the FCRAO Galactic Ring Survey (Jackson et al.\ 2006).
Visually, the similarity between the RJCE extinction and $^{13}$CO emission maps is striking.  
Even small-scale and filamentary features such as the structured cloud between
$25 \lesssim l \lesssim 27^{\circ}$ and above $b \sim 0.5^{\circ}$, or the smaller knot at $(l,b) \sim (20.8, 0.5)^{\circ}$, are prominent in both.
The most significant differences between the maps are where, 
near the $^{13}$CO peak intensities, the RJCE extinction map appears to have ``holes'';
this is precisely the problem discussed in \S\ref{sec:limits}, and one which can be ameliorated by the strategic use of 
IRAC-only maps, at least for these high extinction regions.

The close correspondence between $^{13}$CO emission and high dust extinction is not surprising.  
Both common isotopologues of carbon monoxide, $^{12}$CO and $^{13}$CO, form
in dusty environments, where they are shielded by the dust from dissociation 
in the interstellar radiation field.  The relatively low abundance and higher
excitation threshold of $^{13}$CO make this molecule optically thin even in 
dense molecular clouds, and hence more effective at tracing the total cloud
mass (e.g., Pineda et al.\ 2008).  Thus CO emission, particularly $^{13}$CO, 
is often used as a proxy for dust extinction, provided a dust-to-gas ratio
can be assumed (see work and discussions by, e.g., Bok 1977, Frerking et al.\ 1982, Langer et al.\ 1989, and Dobashi et al.\ 2008).  

The relation between dust extinction in the atomic and molecular phases of the ISM can be used to derive the useful but elusive conversion
factor between CO and H$_2$ emission ($X_{\rm CO}$; e.g., Lombardi, Alves, \& Lada 2006, 
Dobashi et al.\ 2008, Pineda et al.\ 2008, and Liszt, Pety, \& Lucas 2010).  
We will not undertake this analysis here, but future work may enable us to use our
new extinction maps (which in many lines of sight are able to probe to the edge of the Galactic disk) 
to directly explore the relationships among dust, H$_2$, and CO.
Finally, CO emission observations provide useful velocity information that can be 
combined with extinction measures to gauge kinematical distances to dust clouds.

\subsection{Comparison to the IRAC [8.0$\mu$] Emission} \label{sec:rjce-vs-8um}

In Figure~\ref{fig:cp-maps2}d, we show the IRAC [8.0$\mu$] image taken as part of the GLIMPSE-I survey (Benjamin et al.\ 2003).  
This IRAC band contains emission from hot dust
and polycyclic aromatic hydrocarbons, heated by UV radiation in star-forming regions, supernova remnants, or other excited HII regions.  
As such, it is not expected to be a 
particularly effective tracer of the majority of the dust responsible for interstellar extinction, so the clear lack of strong correspondence with the RJCE extinction map
is not worrisome.  We include this panel primarily to help explain discrepancies with the SFD extinction map (\S\ref{sec:rjce-vs-sfd} below) 
and to demonstrate how observations of dust at different wavelengths can reveal quite different and complex behaviors.

\subsection{Comparison to Starcount Maps} \label{sec:rjce-vs-stcnts}

Figure~\ref{fig:cp-maps2}e shows the merged 2MASS+GLIMPSE stellar density in the same region of the Galactic midplane.
Star counts are often used to estimate extinction
in regions where the stellar populations are roughly spatially homogeneous, and where one can depend on a nearby
reference field that is assumed to be relatively reddening-free (e.g., Wolf 1923, Bok 1956, Froebrich et al.\ 2005).  
However, neither of these conditions is generally applicable to the 
Galactic midplane, particularly towards the inner Galaxy.  Nevertheless, we do see reasonable (inverse) correlation between
starcounts and RJCE-derived extinction, particularly in regions of high extinction,
because of the increasing loss of stars behind the denser dust complexes (as in \S\ref{sec:limits}).

This correspondence offers reassurance that our maps are not significantly affected by foreground (low-reddening) dwarf contaminants,
and that even a small number of stars behind a cloud is sufficient to make a reasonable estimate of the line-of-sight extinction 
through the cloud (e.g., $[l,b] \sim [20.8,0.5]^\circ$).
A close comparison of Figures~\ref{fig:cp-maps2}b and \ref{fig:cp-maps2}e also allows one to see where the RJCE extinction map is limited due to a lack of background
stars --- these ``emptier'' patches of sky (e.g., near $[l,b] \sim [23.3,-0.3]^\circ$ or $[28.3, 0.1]^\circ$) show the location of particularly dense cores.

\subsection{Comparison to SFD} \label{sec:rjce-vs-sfd}

The currently most commonly used global reddening maps are those made by Schlegel et
al. (1998; ``SFD'')
 from COBE/DIRBE and IRAS/ISSA data.  These replaced the
previously popular Burstein \& Heiles (1978, 1982) maps made using 
the variation of galaxy counts with position in the sky. 
Low resolution is a primary shortcoming of either survey, particularly at low Galactic latitudes where the
spatial reddening variations can be steep.  The SFD maps have a
$6^\prime$ resolution and are now well-established to become less
reliable with increasing extinction, both at low latitudes and elsewhere.

For example, in a study 
toward the Taurus dark cloud, Arce \& Goodman (1999a) found that the SFD maps
overestimate reddening by 30-50\% when $A(V)$$>$$0.5$, but tend to
{\it underestimate} $E(B$$-$$V)$ in regions with steep extinction gradients.
Chen et al.\ (1999), Stanek (1998a,b), Ivans et al.\ (1999), and von Braun \& Kaspar (2001) 
tested the SFD maps with open and globular clusters and found a range of SFD overestimates
for $A(V)$ by factors of 1.16--1.5.  
Contemporary and subsequent studies using a wide variety of tracers have also derived
significant $V$-band overestimates using the SFD maps: see Cambr\'esy et al.\ (2005), 
Choloniewski \& Valentijn (2003), Rocha-Pinto et al.\ (2004), Dutra et al.\ (2003a,b), 
Yasuda et al.\ (2007), and Am\^ores \& L\'epine (2005, 2007) for examples.
Even SFD themselves noted that a slight trend of residuals existed between their maps and 
reddenings derived from the Mg$_2$--($B-V$) relation,
suggesting that the DIRBE/IRAS maps overestimate the highest reddening
values. 
Chen et al.\ (1999) stressed that there were two crucial simplifications assumed for the SFD maps:
(1) that all of the dust mapped is at a single temperature, and (2) 
that the 40 arcmin beam of DIRBE accurately measured that temperature. 
Although these assumptions are applicable to the low-opacity, translucent cirrus that dominates at high Galactic latitudes,
they are not true for midplane latitudes.  Here, the dust emission is not well-reproduced by a single temperature model,
and it contains significant contributions from small-scale, clumpy molecular gas clouds  (Reach et al.\ 1995; Lagache et al.\ 1998).

The SFD map was created assuming a standard
extinction law, though it is well-known that the interstellar extinction
curve is variable in the optical, with $A(V)/E(B$$-$$V)$ ranging
from 2.6 to 6 for a variety of dust environments (Cardelli et al.\ 
1989); thus it is possible 
to understand some of the shortcomings of the
SFD map given in terms of $E(B-V)$.  
In the map shown in Figure~\ref{fig:cp-maps2}a, we chose not to convert the reddening 
to extinction to avoid adding further uncertainties from the dispersion in 
$A(V)/E(B$$-$$V)$ and $A(K_s)$/$A(V)$.  We have, however, adopted the prescription of
Bonifacio et al.\ (2000) for regions with $E(B-V)$ $>$ 0.1:
\begin{equation}
E(B-V) (>0.1) = 0.1+0.65(E[B-V]_{SFD}-0.1).
\end{equation}

\noindent
The result, shown in Figure~\ref{fig:cp-maps2}a, can be compared to the RJCE-derived map (based on
$(H-[4.5\mu])$ colors) in Figure~\ref{fig:cp-maps2}b.  Of course, the immediately evident
difference is that of resolution, which is much coarser in the SFD map.  
A second striking difference is that the very brightest ``knots" of extinction in the SFD 
extinction map correspond in some cases to extinction ``holes" in the RJCE map.
(We note, however, that the extinction ``hole'' near $(l,b) \sim (28.3, 0.1)^{\circ}$ also appears 
dark even in the IRAC [8.0$\mu$] image (\S\ref{sec:rjce-vs-8um});
this region is spatially associated with several known dark clouds 
(e.g., Carey et al.\ 1998), and the hole at 8$\mu$m indicates 
the presence of dust blanketing so opaque that even IRAC-only data (as in \S\ref{sec:limits}) are too shallow to pierce it.)

Beyond these two differences, comparison of the extinction values of the two maps reveals a broad
correspondence to one another but also notable areas of significant discrepancies.
For example, there are prominent features in the RJCE map that do not appear in the
SFD map, and their absence is not simply due to being blurred out by the lower resolution.
The patchy feature at $(l,b) \sim (26.6, -0.75)^{\circ}$, 
the small but distinct cloud at $(l,b) \sim (20.8, 0.5)^{\circ}$, and
the large swathe of extinction from $25 \lesssim l \lesssim 27^{\circ}$ and above $b \sim 0.5^{\circ}$ 
are examples of the variety of features plainly seen in the RJCE map (Figure~\ref{fig:cp-maps2}b) 
but missing in the SFD map (Figure~\ref{fig:cp-maps2}a),
whereas these same features are clearly visible
in the completely independent $^{13}$CO
map shown in Figure~\ref{fig:cp-maps2}c
(\S\ref{sec:rjce-vs-co}),
demonstrating that these structures are real.
On the other hand, several bright knots in the SFD map {\it not} appearing in the RJCE map 
--- e.g., those at $(l,b)$ =
$(25.4, -0.2)^{\circ}$ and $(20.7, -0.1)^{\circ}$ ---
also do not stand out strongly in the $^{13}$CO map, supporting the notion that 
their non-appearance in the RJCE map is not just a result of stars being extinguished out of these maps.
That the starcount map in Figure~\ref{fig:cp-maps2}e (\S\ref{sec:rjce-vs-stcnts}) shows no special depression in counts at these positions
is further support of this assertion.

Many of the differences between the SFD and RJCE maps may be readily explained by 
comparing Figure~\ref{fig:cp-maps2}a to Figure~\ref{fig:cp-maps2}d, which is the
IRAC [8$\mu$] image of this GLIMPSE region (\S\ref{sec:rjce-vs-8um}) and which bears a striking resemblance to the SFD
map.  The MIR image shows clearly that many of the ``high extinction knots" appearing in the
SFD map are due to the presence of supernova remnants, star-forming regions, and other knots of HII, 
which are strongly affecting the far infrared emission in the DIRBE maps
at these localized positions.
On the other hand, features that are
missing from the SFD map but found in the RJCE maps (e.g., those mentioned above)
 as well as the CO maps (which are
good proxies for the presence of colder dust)
clearly demonstrate that there are strongly light-extinguishing dust clouds
sufficiently cold to have only faint FIR emission, or none at all.

In summary, very close correlations are found among the RJCE extinction map,
the $^{13}$CO emission map, and the 2MASS stellar density map,
the latter two expected to be reasonable tracers of extinction by cold dust (for stellar density, 
only in fairly crowded regions).  In contrast,
significant differences are seen between the quantity and distribution of RJCE extinction and the 
reddening derived from 100$\mu$m dust emission, the
latter which is much more closely correlated with the 8$\mu$m hot dust emission, 
such as that from SN remnants and HII regions.  From these comparisons, we conclude
that RJCE directly and reliably traces interstellar extinction on fine scales.  We also emphasize 
the complexity of dust behavior, 
and the need for studies to consider the suitability of available extinction corrections to the 
observations being corrected.

\section{Some Example Applications} \label{sec:ex-apps}

One goal of this paper is to make the case that the RJCE dereddening technique holds great 
promise as a new tool for Galactic structure (of, in particular, the low-latitude Galaxy), 
interstellar medium, and stellar population studies.
In forthcoming papers, we aim to refine and exploit the methodology for these purposes.

Among the applications we intend for RJCE are (1) continual improvements in the 
systematic two-dimensional mapping of total dust extinction at low
latitudes, by extending the work shown in Figure~\ref{fig:cp-maps2} and offered in our first
generation, public-release maps described in Paper III.
In addition, by exploiting the stellar populations information that is preserved via use of the 
RJCE method, reliable photometric parallaxes can be inferred for large numbers of different
stellar tracers.  This enables one to (2) extend the mapping of Galactic dust into {\it three} dimensions,
and, of course, (3) effectively map the 3-dimensional distribution
of stars across the disk, bulge, and bar of the Galaxy, with the hope of unveiling
in greater detail the shapes and substructure of these dust-enshrouded Milky Way features.
We preview work toward both of these latter goals below.
Finally, it is worth emphasizing again the usefulness of RJCE-dereddened CMDs for identifying
specific types of stars across the inner Galaxy, a desirable feature for assembling homogeneous
samples for further study; this stellar type discrimination of RJCE is an advantage that we are exploiting in, 
e.g., the selection of red giant stars for spectroscopic study via
the new Apache Point Observatory Galactic Evolution Experiment
(APOGEE) as part of the Sloan Digital Sky Survey III (Majewski et al.\ 2010).

\subsection{3-D Extinction Distribution} \label{sec:3dmap}

\begin{figure} 
\begin{center}
\includegraphics[width=0.48\textwidth]{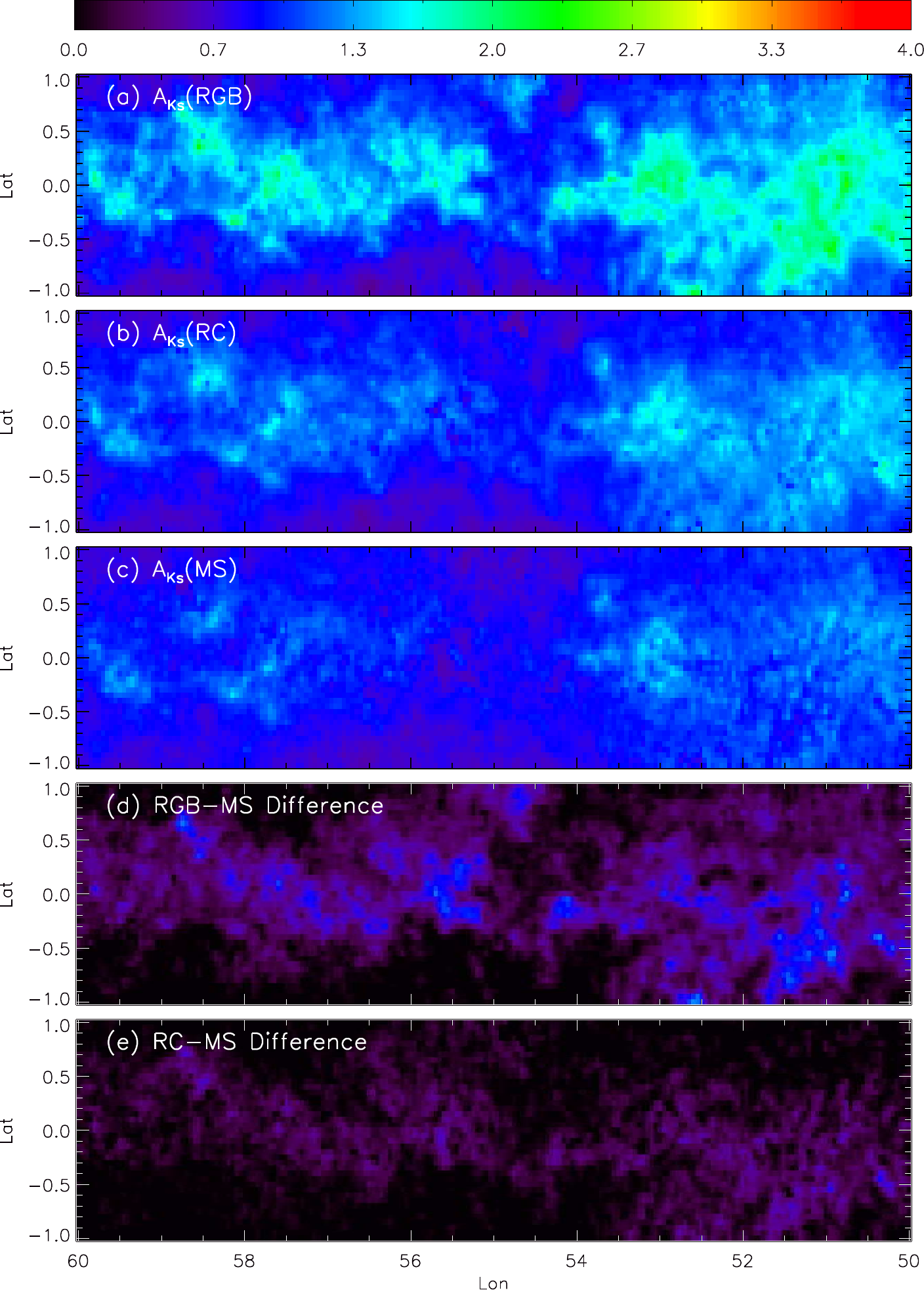}
\end{center}
\caption{RJCE-derived $A(K_s)$ maps of 50$^\circ$ $\leq$ $l$ $\leq$ 60$^\circ$ in the GLIMPSE data,
but using only stars with RJCE-dereddened colors $0.85$ $\leq$ $(J-K_s)_0$ $\leq$ 1.4 to define the red giant branch (RGB, panel a),
$0.55$ $\leq$ $(J-K_s)_0$ $\leq$ 0.85 to define the red clump (RC, panel b),
and $0$ $\leq$ $(J-K_s)_0$ $\leq$ 0.5 to define the main sequence (MS, panel c).  
Panel (d) shows the difference between the RGB and MS maps, and panel (e) the difference between the RC and MS maps.
For each map pixel, we adopt the 90th percentile extinction measured in that pixel.}
\label{fig:pop-maps}
\end{figure}

While the comparisons in \S\ref{sec:cp-maps} suggest that the RJCE methodology performs slightly better than that of the NICE family
for two-dimensional dust mapping, the real advantages of RJCE accrue when one 
considers that this latter method retains sensitivity to stellar type information and luminosity class (\S\ref{sec:cp-cmds}).  
The ability to sort stars as, say, MS versus RC versus RGB type
(via their position in the RJCE-dereddened CMD) allows one to key on stars probing a wide range of
different distances.  As a demonstration of the potential distance sensitivity
and the hope for mapping the three dimensional distribution of dust in the Galaxy, Figure~\ref{fig:pop-maps}
shows RJCE-derived extinction maps spanning $50^\circ$ $\leq$ $l$ $\leq$ $60^\circ$ 
for RGB stars (panel a), less distant RC stars (panel b), and even less distant MS stars (panel c).
To make these maps, all stars in this region (with  {\it JHK$_s$} and [4.5$\mu$] photometric uncertainty $\leq$ 0.5 mag) were
first dereddened using the RJCE method applied to $(H-[4.5\mu])$ colors, and then
the $([J-K_s]_0,[K_s]_0)$ color-magnitude diagram was partitioned into ``MS'', ``RC'', and ``RGB'' divisions by 
appropriate $(J-K_s)_0$ colors.
Comparison of the three maps reveals a variety of cloud structures that lie between the typical distances of the RGB stars
($\sim$18-20 kpc), the RC stars ($\sim$8 kpc), and the MS stars ($\sim$3 kpc) used in these maps. 
For example, the cloud clearly visible in the RGB map at $(l,b)\sim(54.6,0.8)^\circ$ does not appear in the MS map or even the RC map,
which indicates that it is a distant feature beyond the extent of the RC stellar sample.  
Likewise, the filamentary structure centered on $(l,b)\sim(59.0,0.5)^\circ$ apparent in both the RC and RGB maps but not the MS one traces
an intermediate-distance (3--8 kpc) cloud.  An even closer structure appears at $(l,b)\sim(53.7,0.5)^\circ$, 
easily visible in the relatively nearby MS extinction map in addition to the two more distant ones.
We note that this approach to making 3-dimensional dust maps does not rely on assumed
Galactic models, as has been done in previous approaches to this problem (e.g., Marshall et al.\ 2006). 

Obviously, being able to separate stars by their type also enables us to focus on the most
distant, red giant stars, facilitating more reliable maps of {\it integrated} extinction along
a line of sight.  As discussed above (\S\ref{sec:cp-maps}), one of the hindrances to previous attempts to 
map thick dust clouds is the difficulty of ensuring that the stars tracing the cloud are truly behind it.  
The usefulness of relying on the most distant possible tracers is demonstrated in Figure~\ref{fig:pop-maps}
by the fact that the map generated from giant stars highlights much deeper regions of extinction (higher
$A[K_s])$ than the dwarf-star-generated map.

\begin{figure} 
\begin{center}
\includegraphics[width=0.47\textwidth]{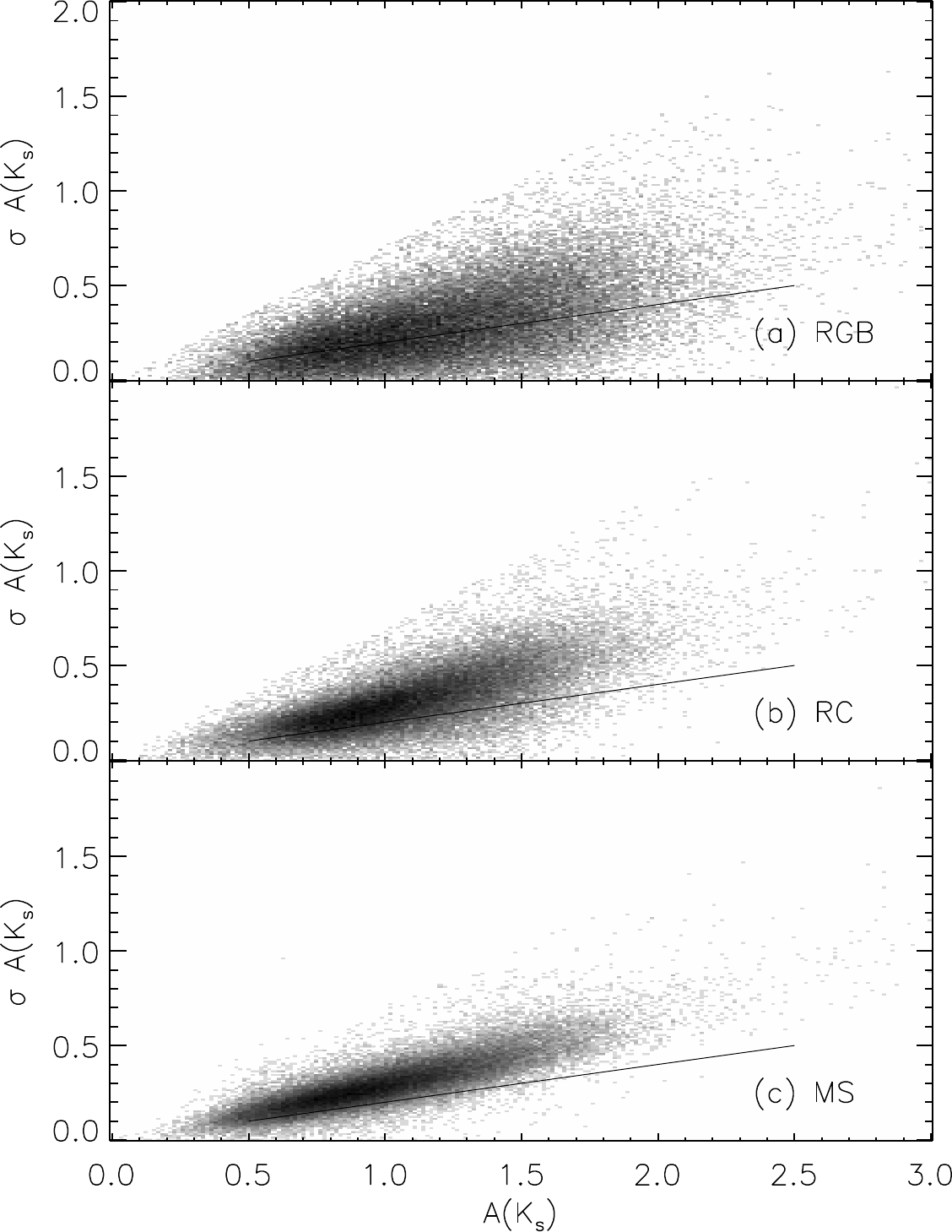}
\end{center}
\caption{Dispersion in $A(K_s)$ extinction per pixel as a function of $A(K_s)$ for the different maps in Figure~\ref{fig:pop-maps}: 
(a) main sequence (MS), (b) red clump (RC), 
and (c) red giant branch (RGB).  The color selection criteria used to defined these populations
are identical to those in Figure~\ref{fig:pop-maps}.
The solid lines are identical to those in Figure~\ref{fig:disp}, to guide the eye in comparison.  
All extinctions were derived with the RJCE method using $(H-[4.5\mu])$ colors.}
\label{fig:pop-disp}
\end{figure}

In addition, if one adopts stellar tracers of only one type, not only are the intrinsic colors more
homogeneous, but so too is the range of stellar distances probed by stars of this type more confined.
We can demonstrate the impact of
these increased homogeneities via the extinction dispersion plots shown in Figure~\ref{fig:pop-disp}, shown separately
for each stellar type used in Figure~\ref{fig:pop-maps} (MS, RC, and RGB).  The net slopes for each of 
these dispersion trends are equal to or lower than in the case of using all stars (compare to Figure~\ref{fig:disp}), 
and we see that the slope of the trends drops as we proceed from MS and RC star tracers (slope
of 0.34) to RGB stars (slope of 0.28).  This trend of decreasing slope
reflects the increasing distances of the dust tracers and therefore the greater likelihood that any
particular tracer will be beyond most of the dust along this line of sight. 

Furthermore, the {\it spread} of dispersion for each tracer is decreased in comparison to that of the all-star sample of Figure~\ref{fig:pop-disp}c, 
because the distance probed by each stellar tracer becomes more uniform.  
In addition, we note a trend between the spread of dispersion and approximate distance of the stellar tracers ---
for example, the relatively close-by MS stars have the tightest spread, because for a fixed angular pixel size, the spatial volume probed
is smaller and therefore less likely to be impacted by sub-pixel-scale dust structures.
Figure~\ref{fig:pop-disp} demonstrates once more that use of predominantly RGB tracers
appears to be the most effective way to improve the reliability of two-dimensional extinction maps
in the Galactic plane.

\subsection{3-D Stellar Distribution} \label{sec:3dstars}

In the previous section we showed, as an example, a general statistical method for gauging the rough
distances to large dust features by taking advantage of
the varying typical distances of stars in maps made by different types of stellar
dust tracers.  
Finer detail in all three dimensions requires greater precision in the stellar distances, which is possible
by estimating photometric parallaxes for individual stars.  The latter is also needed if one wants an accurate
assessment of the distributions of the stars themselves.  Precise photometric parallaxes depend on 
accurately dereddened colors and having a good sense of the type of star (i.e., luminosity class)
one is dealing with.  It has been shown repeatedly here (e.g., \S\ref{sec:cp-cmds} and \ref{sec:3dmap}, and Figure~\ref{fig:pop-maps})
how properly-dereddened
CMDs allow one to discriminate several stellar types
that can serve as very useful standard candles for mapping stellar density distributions.
In Paper III, we will discuss the specific challenges and solutions involved in photometric parallax
determinations for the different types of stellar tracers, but here we address the more general
issue of RJCE map distance limits.

For a certain span of Galactic longitude, GLIMPSE and other similar {\it Spitzer}-IRAC surveys 
contain photometry for RGB stars
all the way to the edge of the MW disk, 
so this population has the potential to provide very useful
constraints on not only the size and shape of the disk but also
the total integrated extinction through the midplane.  
Figure~\ref{fig:sfd_cmd}e demonstrates how we know that GLIMPSE probes 
to the edge of the disk along some lines of sight (in that case, at $l = 307^{\circ}$)
by virtue of the ``edge'' of RGB stars in the CMD (illustrated by the diagonal line
in that panel).  The sudden decline of stars with magnitude at each color is not
due to survey incompleteness, which happens at much fainter magnitudes.  This
RGB edge in the CMD has a slope similar to that of an RGB from a solar metallicity isochrone,
which is precisely the line shown in Figure~\ref{fig:sfd_cmd}e (from Ivanov \& Borissova 2002), but shifted
to an apparent magnitude corresponding to a 16 kpc distance.
At this Galactic longitude, and assuming the Sun is 8.0 kpc from
the Galactic center, the line corresponds to a radius for the Galactic
disk of 12.8 kpc --- i.e., close to the 4 scalelengths one expects for the MW disk.  

\begin{figure} 
\begin{center}
\includegraphics[width=0.47\textwidth]{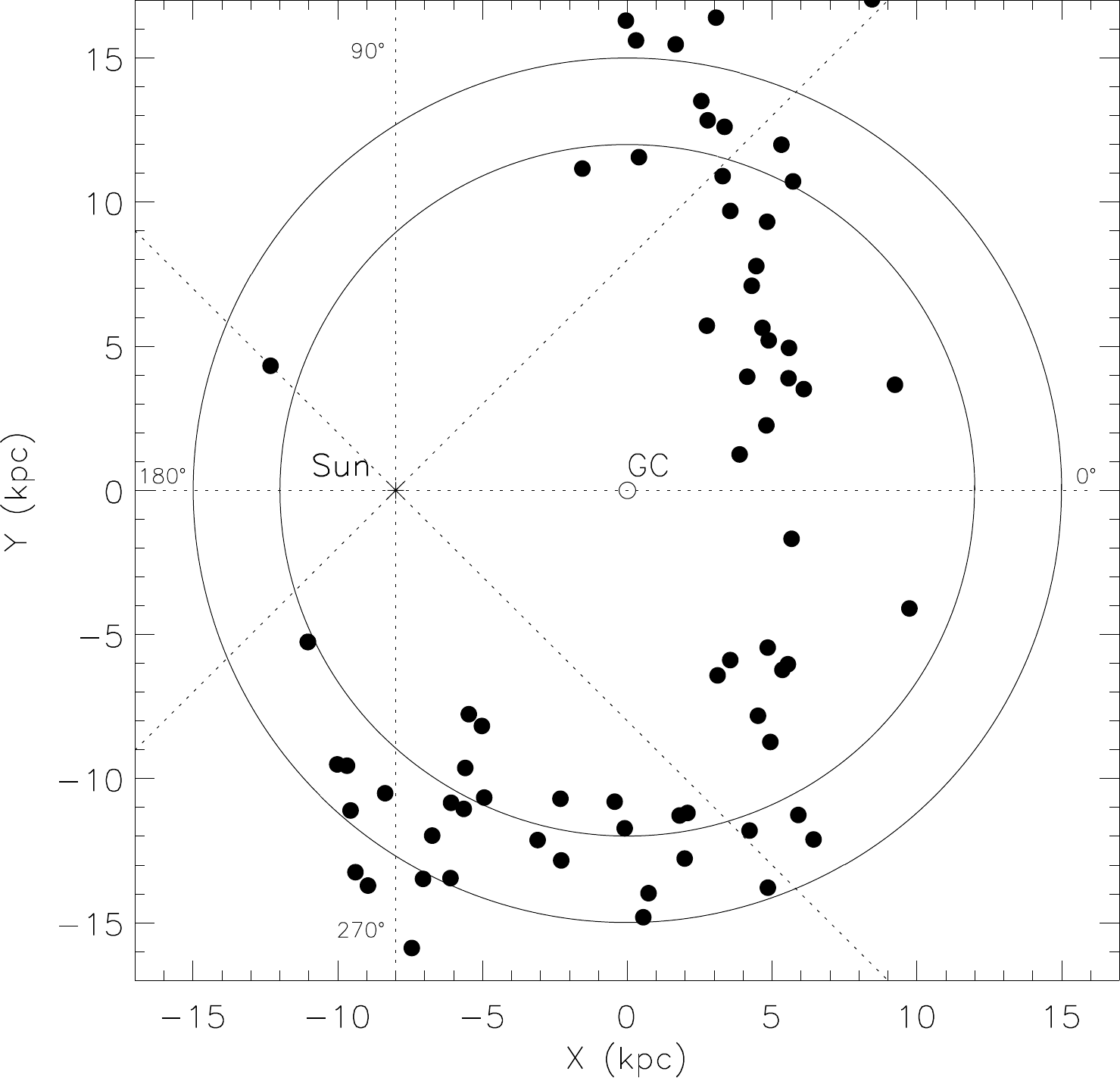}
\end{center}
\caption{Maximum distances probed by NIR CMDs as a function of Galactic longitude, 
as determined from isochrone fits to the lower edge of the dereddened RGB.
The solid circles indicate Galactocentric radii of 12 and 15 kpc.}
\label{fig:diskedge}
\end{figure}

We have undertaken a more systematic and careful analysis of the density drop of the RGB
stars in the CMD across all of the GLIMPSE fields.  For this exercise, we extend the GLIMPSE longitude coverage with an
additional $\sim$40$^\circ$ of outer Galaxy {\it Spitzer}-IRAC data (proposal IDs 40719 and 20499; PI Majewski).
All midplane ($|b|$ $<$ 1.5$^\circ$) stars were dereddened using the RJCE method and then sorted into
longitude bins of $\Delta l$ $=$ 2$^\circ$.  In each bin's CMD, we determined the $K_s$ magnitude of the
faint ``edge'' of the RGB by splitting the RGB into color bins of $\Delta(J-K_s)_0$ $=$ 0.025, fitting
a Gaussian to the $K_s$ distribution in each color bin, and 
then adopting as a standard ``limit'' to the disk's extent the $K_s$ mag that is 1.5$\sigma$ fainter
than the distribution's peak.\footnote{The 1.5$\sigma$ limit was selected as a
repeatable measure of the RGB edge, based on fits ``by eye'' in numerous test fields.}  
The array of color bin centers and RGB edges in each field's CMD were compared to the Padova
suite of RGB isochrones (Girardi et al.\ 2002), leaving distance as a free parameter and using
$\chi^2$-minimization to select the best-fitting isochrone/distance combination.  This method yields an
estimated age, metallicity, and distance for the stars at the faint edge of the RGB (presumably, the most
distant stars) for each longitude bin's CMD.  The age/metallicity grid spanned by the isochrones is relatively
coarse, but we do find subsolar metallicities and ages between $\sim$7--10 Gyr to be the most commonly fit.  Figure~\ref{fig:diskedge}
shows the derived Galactic positions of these most distant stars, for fields fit reasonably well with isochrones (i.e.,
with $\chi^2$ $<$ 3, a quality limit established by visual inspection of the isochrone fits).
One can see that over a large range of longitude (45$^\circ$ $\lesssim$ $l$ $\lesssim$ 315$^\circ$), a common Galactocentric
distance of the disk ``edge'' is found.

\begin{figure} 
\begin{center}
\includegraphics[width=0.33\textwidth,angle=90]{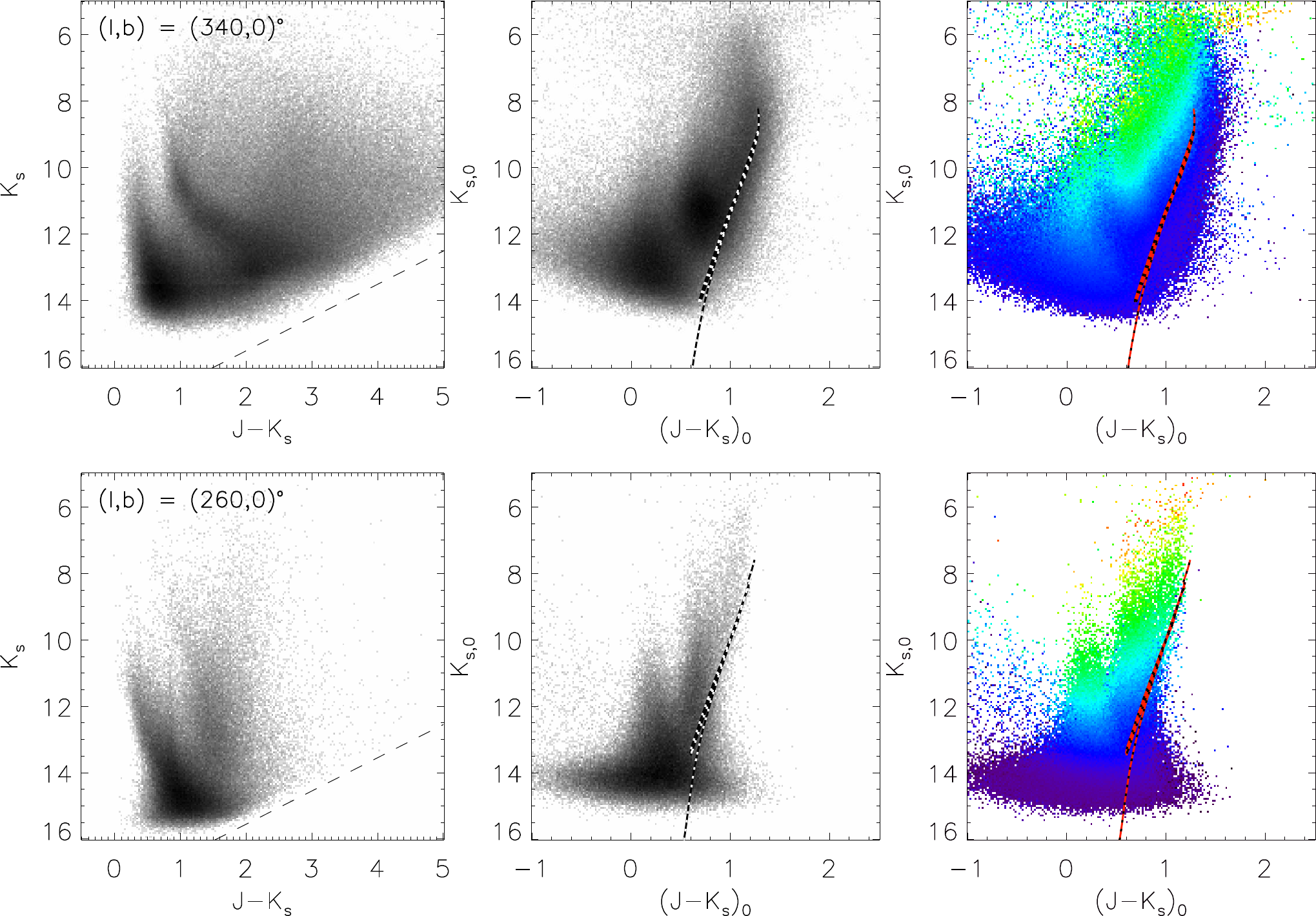}
\end{center}
\caption{{\it Top:} Observed ({\it left}) CMD at ($l,b$) $=$ (340,0)$^\circ$, 
showing the most reddened stars approaching the survey detection limits (dashed line).  The {\it center} and {\it right} panels both show the RJCE-corrected
CMD, but the coloring in the right-hand panel indicates the
average offset of the stars from the survey magnitude limit (dashed line) in uncorrected color-magnitude space, measured along a reddening vector.  
In this scheme, dark blue indicates stars very close to the survey limit, and red indicates stars far removed from the limit.
{\it Bottom:} Same plots as above, but for ($l,b$) $=$ (260,0)$^\circ$, a less-reddened field where the RGB extent is limited by the disk edge only.  
In the dereddened CMDs, the dotted lines are the best-fit isochrones, shifted to the best-fit ``edge'' distances.}
\label{fig:diskedge2}
\end{figure}

Unfortunately, Figure~\ref{fig:diskedge} also clearly demonstrates that even when well-fit by an isochrone, the faint edge of an RGB
may not actually trace a true physical drop-off in stellar density.  The vertical ``wall'' due to near-constant derived heliocentric distance
on the far side of the Galactic Center emphasizes that for $|l|$ $\lesssim$ 45$^\circ$, the distance range spanned by the RGBs
in our CMDs is restricted by the magnitude limits (and possibly crowding effects) of the 2MASS photometry.  One way to identify
the longitudes at which these limits are the dominant factor, without relying on an assumed Galactic disk model, is to measure
how closely the stars at the measured faint edge of the dereddened RGB actually lie to the nominal magnitude limit of the survey.
Figure~\ref{fig:diskedge2} illustrates this point.  In the top
row of panels, with midplane stars at $l$ $\sim$ 340$^\circ$,
the lower edge of the stellar distribution in the left-hand panel follows closely the dashed line indicating the approximate survey detection
limits\footnote{http://www.ipac.caltech.edu/2mass/overview/about2mass.html}.  (The limits shown here are for ideal, unconfused fields; 
the source confusion present in low latitude fields may account for the actually observed, brighter limit, 
but note that the slope of the limit is identical to the apparent cutoff slope of the data.)  When
the CMD is dereddened and an isochrone fitted, the stars being fit at the edge of the RGB are largely those lying near the 
detection limit (denoted by dark blue colors in the right-hand panel); this indicates that while the isochrone may in fact provide
a reasonable estimate of the age/metallicity/distance of the stars, those stars are not probing the physical edge of
the Galactic disk.  As a contrast, we show in the bottom panels the same
CMDs for a midplane field at $l$ $=$260$^\circ$;
even in the uncorrected CMD, it is clear that the majority of RGB stars at the faintest magnitudes for each color
do not have apparent brightnesses approaching the survey limit (confirmed by the colors of those stars in the lower right panel), so
we may be confident that in this field, the fitted isochrone is actually measuring stars at the outermost extent of the disk.  

The most important implications of this simple isochrone fitting procedure are twofold.  First, we are able to
indicate at which longitudes our integrated
extinction maps likely explore the actual total Galactic extinction (from Figure~\ref{fig:diskedge}, 
we assess this to be $|l|\gtrsim45^{\circ}$),
as well as the approximate distances probed by the inner Galaxy maps.  Second, the isochrones fitting those RGBs that
{\it do} sample the edge of the Galactic disk can
provide a measurement of the stellar characteristics and distances of the edge of the Galaxy.  We will
explore and refine this method in future work, 
but even this simple interpretation of the CMD in terms of the Galactic distribution of disk stars
demonstrates the potential of systematically exploring properly dereddened CMDs
for a better understanding of the Galactic midplane's stellar populations.

\section{Conclusions and Future Work} \label{sec:conc}

We approach the well-tested method of using IR color excesses to calculate foreground dust extinction from a new angle
--- we combine near- and mid-infrared photometry (2MASS + {\it Spitzer}-IRAC) to trace reddening in a way far less 
dependent on assumptions of stellar type than previous studies have been.  
This Rayleigh-Jeans Color Excess (RJCE) technique is a major improvement over previous variants
because it samples starlight on the Rayleigh-Jeans portion
of the stellar spectral energy distribution (SED),
where the vast majority of stellar SEDs have a common shape.
Thus, our method is far less susceptible to variations in intrinsic stellar colors, so that
measured color excess are much better matches to the true reddening of the star.
The RJCE method makes it possible to measure rather accurately the amount of selective and total
extinction foreground to any particular star.

One of the most significant improvements made by RJCE is that the reddening corrections it provides
are virtually completely independent of the intrinsic color for most stars.  Thus, application of
the method preserves intrinsic stellar colors and allows restoration of the {\it intrinsic} color-magnitude
diagram with high fidelity --- in essence, allowing recovery of the intrinsic stellar types through photometry alone.
We show that using this additional step of the RJCE method in heavily-reddened
midplane fields produces a NIR color-magnitude distribution extraordinarily similar to that 
predicted by a zero-reddening Galactic model; 
in contrast, the stellar type assumptions required by other color excess methods based on
only NIR photometry (the NICE and NICER methods)
result in unphysical color-magnitude distributions and a loss of stellar type information.
The availability of color-magnitude diagrams free of reddening for even very dusty, 
Galactic midplane lines of sight with highly variable extinction
opens up unprecedented opportunities for stellar population studies of the inner, low-latitude Galaxy.

The RJCE-generated extinction maps are compared with numerous other proxies for dust extinction: 
the 100$\mu$m $E(B-V)$ map by Schlegel, Finkbeiner, \& Davis (1998; SFD), 
the $^{13}$CO emission map from the Galactic Ring Survey (Jackson et al.\ 2006), 
the 8$\mu$m dust emission image from the GLIMPSE program using {\it Spitzer}-IRAC (Benjamin et al.\ 2003), 
and the 2MASS star counts of the same region.  
We find poor correlation with the $E(B-V)$ map of SFD.  This is because of the lower resolution
of the SFD maps, strong variations in the optical extinction law, the fact that the SFD maps represent total line-of-sight extinction (not extinction foreground
to each star), and, as has been demonstrated
before by previous authors, that the SFD maps overestimate the reddening at low latitudes.
Most importantly, the poor correlation between SFD and RJCE maps
shows that infrared emitting dust, which provides the basis of the SFD extinction maps,
is not a good proxy for infrared extinguishing dust.

On the other hand, we find an excellent match, down to even 
fine filamentary features, between RJCE-generated $A_K$ maps and maps of the molecular $^{13}$CO 
emission.  Clearly the $^{13}$CO molecules and light-extinguishing dust grains are closely linked; in future work we
aim to explore this correlation more quantitatively.  This striking correspondence between
the $^{13}$CO and RJCE-generated $A_K$ maps confirms that the latter are able to 
trace light-extinguishing dust on very small angular scales and
that RJCE maps should provide a more accurate measurement of the interstellar extinction in the Galactic midplane than the SFD maps.
To this end, we provide in Paper III a Version 1 release of extinction maps for those regions covered by
the GLIMPSE-I and Vela-Carina surveys, using the $(H-[4.5\mu])$-based RJCE techniques described here.  
We intend to extend and improve these maps as our program evolves.

We also explore the photometric limitations of the 2MASS+IRAC catalog
and the effect of these limitations on our derived extinction maps, especially in 
the most highly extinguished lines of sight.  It is demonstrated how,
with either deeper NIR photometry or exclusive use of MIR photometry, this generally
minor drawback (in terms of fraction of the sky affected) can be overcome.
We explore the various NIR+MIR color combinations to identify those
expected to make the RJCE technique the most effective, and we
discuss the advantages and disadvantages of single- versus hybrid-color schemes
in its application.

Among these opportunities provided by the RJCE method's ``second step'' --- the recovery of intrinsic stellar types --- 
is the ability to cleanly discriminate various types of
stellar population tracers.  For example, in low latitude NIR CMDs three main stellar types are 
most evident: main sequence, red clump and red giant branch stars.  RJCE-dereddening allows
relatively pure samples of these tracers to be isolated, and, to the extent that these stellar types
are standard candles, enables the mapping of both dust and stellar distributions in three dimensions.
We preview this exciting new capability by showing extinction maps generated independently for 
these specific stellar populations --- each of which probes distinctly different distance ranges for
2MASS+IRAC data --- and finding examples of cloud formations that can be isolated to specific
distances.  
Maps of this type for the full GLIMPSE-I and Vela-Carina survey areas are described in Paper III.
As a second example, the 2MASS+IRAC CMDs are analyzed over a wide longitude range, 
with many showing strong evidence that their RGBs sample the outer edge of the Milky Way stellar disk; we measure
this edge for $ | l | \gtrsim 45^{\circ}$ and find it to be consistent with a Galactocentric distance of 12--15 kpc, or $\sim$4--5 
disk radial scalelengths.

There are many opportunities for application and improvement of the RJCE methodology,
and a broad program of experiments is planned.
One is to improve the RJCE extinction estimates using a more refined treatment of stellar properties; 
this will be aided by the upcoming APOGEE survey, which will allow us to compare RJCE-derived 
stellar types to spectroscopically-determined ones.  
We are evaluating the benefit of of optimized ``hybrid''-color schemes where high extinction limits the usefulness of NIR photometry.
We have already taken advantage of our
data set to measure the extinction behavior as a function of wavelength throughout the midplane (Paper II), and
work is under way to provide RJCE extinction maps for public use
(Paper III, in preparation) and to use the new, reliably-cleaned CMDs to explore Galactic structure in previously-hidden segments of the Galaxy.

\begin{acknowledgments}
We thank R.~Indebetouw and M.~F.~Skrutskie for helpful discussions, M.~Meade and B.~Babler for their efforts to reduce the Vela-Carina dataset
using the GLIMPSE pipeline, and J.~K.~Carlberg and the anonymous referee for comments and advice that have greatly 
improved the clarity of the paper.
SRM appreciates the hospitality of the Observatories of the Carnegie Institution of Washington for hosting a sabbatical visit during which
this project was conceived.
We acknowledge generous support from The F. H. Levinson
Fund of the Peninsula Foundation, NSF grants AST-0307851
and AST-0807845, and funding from NASA {\it Spitzer} grants 1276756 and 1316912.
DLN has been supported by an ARCS Scholarship, a VSGC Graduate Research Fellowship, and an SDSS-III APOGEE postdoc, and
GZ has been supported by a VSGC Graduate Research Fellowship and a NASA Earth \& Space Science Fellowship.
This work is based in part on observations made with the {\it Spitzer} Space Telescope, and has made use of the NASA~/~IPAC Infrared Science Archive, 
which are operated by the Jet Propulsion Laboratory, California Institute of Technology under a contract with NASA. 
We acknowledge use of data products from the Two Micron All Sky Survey, a joint project of the University of Massachusetts and the 
Infrared Processing and Analysis Center~/~California Institute of Technology, funded by NASA and the NSF.  
This publication also makes use of molecular line data from the Boston University-FCRAO Galactic Ring Survey (GRS). 
The GRS is a joint project of Boston University and Five College Radio Astronomy Observatory, 
funded by the National Science Foundation under grants AST-9800334, AST-0098562, \& AST-0100793.  
\end{acknowledgments}

\end{document}